\documentstyle[11pt]{article}
\title{
{\normalsize
\begin{flushright}
DAMTP--94--21\\
VAND--TH--94--4 \\
astro-ph/9405028 \\
May 1994 \\
\end{flushright}}
\vspace{1 cm}
Anharmonic Evolution of the Cosmic Axion Density Spectrum}
\author{Karl Strobl\\
Department of Applied Mathematics and Theoretical Physics\\
Cambridge University, Cambridge, UK
\and
Thomas J. Weiler\\
Department of Physics \& Astronomy\\
Vanderbilt University, Nashville, TN 37235, USA}
\date{May 4,1994}
\begin{document}
\maketitle

\begin{abstract}
We present analytic solutions to the spatially homogeneous axion field
equation, using a model potential which strongly resembles the
standard anharmonic $(1-\cos N\theta)$ potential,
but contains only a piece-wise second order term.
Our exactly soluble model for $\theta(t)$ spans the entire range
$[-\pi/N,\pi/N]$.
In particular, we are able to confirm
(i) Turner's numeric correction factors \cite{Turner}
to the adiabatic and harmonic analytic treatments of
homogeneous axion oscillations, and
(ii) Lyth's estimate \cite{Lyth} valid near the metastable misalignment angle
$\pi/N$ at the peak of the potential.
We compute the enhancement of axion density fluctuations that occurs
when the axion mass becomes significant at $T\sim 1$ GeV.
We find that the anharmonicity amplifies density \mbox{f}luctuations,
but only significantly for relatively large
initial misalignment angles.
The enhancement factor is
$\sim$ (2,3,4,13) for $\theta_{\rm in}\sim (0.85,0.90,0.95,0.99)\times\pi$.
\end{abstract}
\newpage

\section{Introduction and model}

Axions are a consequence of the simplest solution to the strong CP problem
\cite{PQ}. The complex Peccei-Quinn field $\psi$
has at temperatures above the QCD scale $\Lambda_{\mbox{\scriptsize QCD}}$
an effective potential of the ``Mexican hat" shape.
Thus, $\psi=\rho$e$^{i\theta}/\sqrt{2}\,$, with the classical value of
$\rho\sim$ the Peccei--Quinn breaking scale $f_{\rm a}\,$.
The axion field $a(\vec{x},t)$
corresponds to the angular variable $f_{\rm a}\,\theta$
in the Peccei-Quinn field.
Ignoring variations of $\rho$ in the PQ field (which is reasonable
for temperatures far below the PQ and in\mbox{f}lation scales
\footnote{Models with changing
$\rho$ during inflation have been constructed \cite{delta}. If they are true,
they reopen the axion window to a wider cosmologically allowed range of
axion masses. The calculations presented in this paper, however,
are independent of the inflationary mechanism.})
we get the field equation
\[
\Box \theta(\vec{x},t)+V'(\theta(\vec{x},T(t)))=0\mbox\ .
\]
The effective potential $V$ for the $\theta$--direction is
flat for high temperatures, and the axion is
massless at high temperatures.  At low temperatures
$T\sim \rm{few}\times \Lambda_{\mbox{\scriptsize QCD}} \sim$ 1 GeV,
the azimuthal part of the Mexican hat potential is
tipped ($N=1$) or rippled ($N\ge2$)
due to instanton effects \cite{GPY}, where $N$ is the color anomaly of the
PQ symmetry.  The resulting curvature about the minimum
is interpretable as the induced axion mass squared.
The deviation of the potential from quadratic $\theta$--dependence at larger
$\theta$ is the source of anharmonic effects.  We will present a model
that closely mimics the instanton--induced potential.  Then we will
analytically solve the model to obtain the time--evolution of fluctuations
in the axion field $\theta$, and ultimately, of fluctuations in the axion
density.

In principle,
the initial axion perturbations could arise from stochastic values assigned to
each acausal volume in the early, non--inflationary universe,
or from quantum mechanical field fluctuations
within the causally connected horizon of the inflationary universe.
In the acausal universe, one expects on average one
domain wall per causal volume.
These domain walls are disallowed by the observed smoothness of the universe's
energy distribution, and so the inflationary origin of the axion density
perturbations is preferred.
However, the nature of the mechanism generating the initial fluctuations
does not enter into our calculation to follow.
This is because the appearance of the instanton--induced
effective potential for the axion field occurs at $T\sim$ 1 GeV,
long after the epochs of inflation and PQ symmetry breaking.

There is a subtlety to discuss.
We will work with the homogeneous field equation, i.\ e.\ we will neglect
gradient terms.  A field inhomogeneity exists in a finite spatial region,
and therefore the field has a nonzero gradient.  The gradient will be
negligible for small fluctuations, or for
those fluctuations whose wavelengths are superhorizon during the time when the
anharmonic effects come to dominate the expansion term, at $T\sim 1$ GeV.
It is the evolution of these fluctuations that we consider.
And it is these fluctuations that truly have no memory of their origin,
e.g. whether they are entering (non--inflation) or re--entering (inflation)
the horizon.  Thus, as with all prior work, our analysis is strictly valid
for small fluctuations satisfying $\delta\theta \ll \theta$.
In fact, small scalar field fluctuations are expected in the
inflationary scenario \cite{tilde}.  Thus, axion fluctuations should be small
but nonzero if the universe underwent an inflationary epoch that ended
after $T\sim f_{\rm a}$ and before $T\sim \Lambda_{\mbox{\scriptsize QCD}}$.
Also, the potential domain wall problem which arises when
$\theta=\overline{\theta}(1+\delta\theta/\overline{\theta})$
approaches $\pm\pi$ is mitigated in the small fluctuation universe.
Recently, Kolb and Tkachev \cite{KT} have initiated numerical integration of
the partial differential equations
which include the gradient terms; their results show that when gradients
cannot be neglected, their effects may be dramatic, and that when
fluctuations are large, domain walls will form.

Perturbative calculations of the instanton--origin of the axion mass
are only valid at temperatures $T\gg \Lambda_{\mbox{\scriptsize QCD}}$.
In Turner's work \cite{Turner} a power law
\begin{equation}
m_{\rm a}\propto T^{-n}\mbox\
\label{eq:m(T)}
\end{equation}
is fit to the perturbative instanton result for the potential's curvature
at its minimum; values of $n$ between 3.6 and 3.8 result.
Since the perturbative treatment is not expected to be valid at just the
temperature $T \sim \Lambda_{\mbox{\scriptsize QCD}}$ where the mass is turning
on,
we will follow Turner and study a larger possible range for the
power--law exponent $n$.  Ultimately, the axion mass reaches its
zero--temperature value and ceases to grow.

$\theta$ parameterizes the different QCD vacua, whose
energy densities, neglecting fermions\cite{Coleman}, are given by
\begin{equation}
V(\theta)=K(1-\cos N\theta)\, \mbox{e}^{-S_{\rm o}} \mbox\ , \label{eq:Vcos}
\end{equation}
where $S_{\mbox{\scriptsize o}}$ is the action of a QCD instanton, and the
parameter $K$ is given by
\begin{equation}
K=\left(m_{a} f_{a} /N\right)^2=\left(
f_{\pi} m_{\pi} \frac{\sqrt{z}}{1+z}\right)^2
\simeq(\mbox{79~\mbox{MeV}})^4\mbox\ .
\label{eq:m0}
\end{equation}
The second equality in Eq.\ (\ref{eq:m0}) is a current algebra result
\cite{BarTye}
, relating the pseudo--Goldstone axion to the pseudo--Goldstone pion and to
$z$, the ratio of the up quark to down quark mass.
{}From it one finds, taking $z\simeq 0.5$ and $N=1$,
$(m_{\rm a}/10^{-5}\mbox{eV}) \simeq (f_{\rm a}/10^{12}\mbox{GeV})^{-1}$.
Astrophysical
arguments \cite{axgen} constrain the axion mass to lie in a rather narrow
preferred ``window", $10^{-5}{\rm eV}\stackrel{<}{~} m_{\rm a} \stackrel{<}{~}
10^{-3}{\rm eV}$.  Further argument relates the present cosmic
axion mass density (in units of the closure density $\Omega_{\rm crit}$)
to the axion mass: $\Omega_{\rm a} = \left(\frac{m_{\rm a}}{10^{-5}{\rm
eV}}\right)^{-7/6}$;
an $m_{\rm a}=10^{-5}$ eV axion closes the universe.
The functional form of Eq.\ (\ref{eq:Vcos}) is sensitive to the inclusion
of fermions in the theory, although the periodicity
of the potential is not \cite{Oxford}.
Here we input a functional form that
strongly resembles the familiar form of Eq.\ (\ref{eq:Vcos}),
preserves the periodicity, but
at the same time yields a homogeneous equation of motion which enables
us to obtain an analytical solution for the field.
We accomplish this by matching a parabola to an inverted parabola:
\begin{equation}
V(\theta)= \left\{
     \begin{array}{ll}
     \frac{1}{2}m_{a}^{2}(T)\theta^2, &
\left|\theta\right|<\frac{\pi}{2}\\[2mm]
     \frac{1}{2}m_{a}^{2}(T)\left(\frac{\pi^2}{2}-
     \left(\pi-\left|\theta\right|\right)^2\right), & \left|
     \theta\right|>\frac{\pi}{2}\mbox\ .
     \end{array}
\right. \label{eq:modelV}
\end{equation}
For convenience, we have assumed in Eq.\ (\ref{eq:modelV})
that the color anomaly parameter
$N$ in Eqs.\ (\ref{eq:Vcos}) and (\ref{eq:m0})
is one; our results extend trivially to the $N\geq 2$ cases.
Our model potential is compared to the cosine potential in Fig.\ 1.

Previous analytic work on the evolution of the axion field was restricted to
either the harmonic potential near the minimum at $\theta=0$
\cite{classical axion,Dine},
or near the maximum at $\left|\theta\right|=\pi$ \cite{Lyth}.  In most
cases, an adiabatic aproximation $\frac{dm}{dt}\ll m^2$ was imposed to
facilitate a simple analytic solution \cite{classical axion,Lyth}.
Turner \cite{Turner} numerically
integrated the equation of motion in the harmonic potential
to test the adiabatic approximation, and in the anharmonic cosine potential
of Eq.\ (\ref{eq:Vcos})
for $\left|\theta\right|$ not too near $\pi$,
to test the harmonic approximation.
He found significant corrections.
With our model potential, we are able to derive analytical solutions without
succumbing to the harmonic or adiabatic approximations of prior work.
Our solution thus improves and relates the prior work.

We solve the field equations classically,
which is valid during the epoch of perturbation growth,
long after any possible inflation \cite{Pi}.  The only significant
growth in density perturbations occurs at the time when the
growing axion mass is comparable in magnitude to the diminishing
Hubble parameter \cite{Turner}.
At earlier times the fluctuations are frozen by the overdamped equation of
motion; and for later times, both classical \cite{classical axion,Dine}
and quantum \cite{box} calculations agree that fluctuation growth yields
small overdensities.  As we will show later in Fig.\ 4,
the latter result derives from the fact that the
late--time field has evolved into the harmonic region of the potential,
where fluctuation growth does not occur.

Once $\theta$ is calculated, the energy density per physical volume is obtained
from $T_{\mu\nu}=f_{\rm a}^2
\partial_{\mu}\theta\partial_{\nu}\theta-g_{\mu\nu}
{\cal L}$ as $\rho=T_{00}=\frac{1}{2}f_{\rm a}^2\dot{\theta}^2+V(\theta)$,
ignoring the gradient term.
Eventually, the axion field evolves to small--amplitude, quasi--periodic
oscillations in the harmonic region about the potential minimum.
Then, to a good approximation, $\rho$ is just
$V(\theta_{\rm out})\approx \frac{1}{2}f_{\rm a}^2 m_{\rm a}^2 \theta_{\rm
out}^2$,
where $\theta_{\rm out}$ is the oscillation amplitude and $m_{\rm a}$ is the
zero--temperature axion mass.

If the initial \mbox{f}luctuations are sufficiently small,
we can disregard spatial derivatives in the field equation
and consider only the time--derivative part of the
Klein-Gordon operator $\Box$.
In a radiation-dominated FRW universe, this yields the usual equation of motion
for the homogeneous ``zero--mode'' axion field ($a(t)=f_{\rm a}\theta(t)$):
\begin{equation}
\ddot{\theta}(t)+\frac{3}{2t}\dot{\theta}(t)+m_{\rm a}^2(T(t))\theta(t)=0\mbox{
,}
\label{eq:eom}
\end{equation}
if $\left|\theta\right|\leq\pi/2$, and the same equation with $\theta$ replaced
by $\epsilon=\pm\pi-\theta$
and $m_{\rm a}^2$ by $-m_{\rm a}^2$ for $|\theta|>\pi/2$.
Here the Hubble parameter, $H(t)$, in the second term has been equated to
$1/2t$, the value appropriate for a radiation--dominated FRW universe.

Since the equation of motion is linear and homogeneous in $\theta$ for
$\left|\theta\right| \leq\pi/2$, the mean field $\overline{\theta}$
and the local field $\theta=\overline{\theta}+\delta\theta$ evolve with
the same time--dependence. Thus, the field contrast $\frac{\delta\theta}
{\overline{\theta}}$ is time--independent, i.e. does not evolve.
The same conclusion follows for $\frac{\delta\epsilon}{\overline{\epsilon}}$
at  $|\theta|>\pi/2$.
Thus, in our model, the evolutionary enhancement of
$\frac{\delta\theta}{\overline{\theta}}$
comes from the matching of solutions at $\theta=\pm\pi/2$.
Growth occurs because $\theta=\overline{\theta}+\delta\theta$ must be matched
to $\pm\pi/2$ at a time differing slightly from the matching time of
$\overline{\theta}$.

The argument of the mass in Eq.\ (\ref{eq:eom}) requires a relation
between time and temperature.
In the radiation--dominated FRW universe, the two are inversely
related by $t \propto T^{-2}$.  The relation is \cite{KoTur}
$t^{-1}=3.3\sqrt{g_*}\frac{T^2}{M_P}$, where $M_P$ is the Planck mass
$\sim 1.2\times 10^{19}$ GeV and $g_*$ is the effective number of degrees
of freedom (with $7/8$th's weighting for fermions)
at temperature $T$.  Counting photons, gluons, three neutrinos, two
leptons, and three quark flavors, $g_* = 62$ and $t^{-1}=26 \frac{T^2}{M_P}$.
At $T\!\sim1$~GeV, the value of $3H$ is $\sim 3\times10^{-9}$~eV, which is
much less than the
zero--temperature axion mass.  Thus, once the instanton--induced potential
turns on at $T\sim1$~GeV, its curvature quickly dominates the Hubble
term in the equation of motion.

It is convenient
\footnote{
The easiest way to see that scaling time to $\tilde{t}$ is merely a convenience
is to replace RW time in the equation of motion (\ref{eq:eom}) with
conformal time $\tau(t) \equiv \int^t \frac{d\,t'}{R(t')}$
(so that $d\,\tau=d\,t/R(t)$),
and then to rescale the solution as $\tilde{\chi}\equiv R(\tau) \chi$.
$R(\tau)$ is the scale factor of the conformal metric,
which in the radiation dominated epoch equals $R_{\rm o}\frac{\tau}{\tau_{\rm
o}}$.
The resulting equation of motion then contains no explicit expansion term:
$$
\ddot{\tilde{\chi}} \pm \left[R(\tau)m_{\rm a}(T(\tau))\right]^2 \tilde{\chi} =
0 .
$$
Here, an overdot is a derivative with respect to conformal time.
Since $T$ scales as $1/R$, we have, using Eq.\ (\ref{eq:m(T)}),
$\left[R(\tau)\,m_{\rm a}(T(\tau))\right]^2 =
\left[R_{\rm o}\,m_{\rm a}(\tau_{\rm o})\right]^2
(\frac{\tau}{\tau_{\rm o}})^{2n+2}$,
which makes it clear that any reference point chosen
during the power--law growth period of the mass is equally valid.
The solution to this e.o.m.\ is just
$\tilde{\chi}(\tau)=\sqrt{\frac{\tau}{\tau_{\rm o}}}\,Z_{\pm\frac{1}{2n+4}}
\left(\frac{R_{\rm o} \tau_{\rm o} m_{\rm a}(\tau_{\rm o})}{n+2}
(\frac{\tau}{\tau_{\rm o}})^{n+2}\right)$,
where $Z$ is any Bessel function for the convex potential, and any
modified Bessel function for the concave potential.
Connection with our RW--time solution is made by substituting
the radiation dominance relation $\tau/\tau_{\rm o}=(t/t_{\rm o})^{1/2}$ into
$\chi=\tilde{\chi}/R$ to get $\chi= (t_{\rm o}/t)^{1/4}Z_{\pm\frac{1}{2n+4}}
\left(\frac{R_{\rm o} \tau_{\rm o} m_{\rm a}(\tau_{\rm o})}{n+2}
(\frac{t}{t_{\rm o}})^{\frac{n+2}{2}}\right)$.
We see that our choice made in the text for the unit of time is just a
particular value for the arbitrary reference point
$R_{\rm o} \tau_{\rm o} m_{\rm a}(\tau_{\rm o})$.}
to measure all time in units of $\tilde{t}$,
the time (roughly) when the axion oscillations
for the convex part of the potential first become undamped.
$\tilde{t}$ is defined implicitly by
$m_{\rm a}(\tilde{t})=\frac{3}{2\tilde{t}}$,
which yields $\tilde{T}\sim 1$ GeV and $\tilde{t}\sim 10^{-6}$ sec.
In terms of this scaled, dimensionless time, the field equation becomes simply
\begin{equation}
\ddot{\chi}(t)+\frac{3}{2t}\dot{\chi}(t)\pm\frac{9}{4}t^n\chi(t)=0\mbox{ .}
\label{eq:geneq}
\end{equation}
In the convex part of the effective potential, $\chi$ denotes $\theta$ and the
plus sign holds for the last term;
in the concave part of the potential,
$\chi$ denotes $\epsilon=\pm\pi-\theta$ and the minus sign holds.
The growth of the axion mass terminates when the mass
attains its low temperature value $m_{\rm a}(\infty)$ at $t_c\sim
(m_{\rm a}(\infty)/m_{\rm a}(\tilde{t}))^{2/n}$; or equivalently,
$\sim (2/3 m_{\rm a}^{-1}(\infty))^{2/n}$
with $t_c$ and $m_{\rm a}^{-1}(\infty)$ measured in units of $\tilde{t}$.

If the initial value for $|\theta|$ exceeds $\pi/2$, then each time the
field amplitude passes through $\pm\pi/2$
from above or below, there will be matching
conditions to join together the $\theta$ and $\epsilon$ solutions
extracted from Eq.\ (\ref{eq:geneq}).  The solutions to Eq.\ (\ref{eq:geneq})
are the Bessel and modified Bessel functions, respectively:
\begin{eqnarray}
\theta(t) & = & t^{-\frac{1}{4}}J_{\pm\frac{1}{2n+4}}\left(
\frac{3\,t^{\frac{n+2}{2}}}{n+2}\right)
\label{eq:solution.a}\\*[2mm]
\epsilon(t) & = & t^{-\frac{1}{4}}I_{\pm\frac{1}{2n+4}}\left(
\frac{3\,t^{\frac{n+2}{2}}}{n+2}\right)\mbox\ ,
\label{eq:solution.b}
\end{eqnarray}
To match the initial condition at $t\ll 1$, we will
disregard the Bessel functions with negative order,
since their singular behavior at $t=0$ implies that even a large component
of this solution at the initial time $t_{\rm o}\ll 1$
quickly falls to insignificance by $t=1$.
However, whenever $\theta$ evolves through $\pm\pi/2$ from above or below
at a later stage,
we have to match to the pair of independent solutions given above.
The negative order solutions are the analogues of the singular $1/\sqrt{t}$
mode arising (and customarily rejected) for the early
epoch massless field discussed below.

It is worthwhile at this point to
examine the solution for the convex part of the
potential, and to relate it to previous
work done on homogeneous scalar field evolution.

\section{Mass turn--on with a harmonic potential}

The case of axion oscillations within the regime of a harmonic
potential $V\propto \theta^2$ has been discussed extensively in the literature
\cite{classical axion,Dine}.
The following is a brief review of conclusions from these references.
Before the axion oscillations become underdamped (i.e.\ before $\tilde{t}$),
the mass term is negligible and the solutions to the field equation
are simply $\theta = constant$, and $\theta \propto t^{-1/2} \propto T$.
The constant solution is assumed to dominate when $T$ cools to $\sim 1$GeV.
After the mass exceeds a few times the Hubble parameter, one takes
the axion field to oscillate with frequency $m(t)$,
valid in the ``adiabatic" approximation that
$\dot{m}\ll m^2$.

With these approximations, the ansatz
\begin{equation}
\theta(t)=A(t)\cos(m(t)t)\mbox\ ,
\label{eq:ansatz}
\end{equation}
inserted into Eq.\ (\ref{eq:eom})
yields $A(t)\propto R^{-3/2}m^{-1/2} \propto t^{-\frac{n+3}{4}}$
($R$ being the RW scale factor).
With this adiabatic ansatz, the dependence of the axionic mass density
on the initial misalignment angle is easily calculated.
Apparently, $m A^2 R^3$ is an adiabatic invariant; call it $J_A/f^2_{\rm a}$.
Since the axions are nonrelativistic, we see that the axion energy density is
$\rho_{\rm a}=\frac{1}{2} m_{\rm a}^2 f_{\rm a}^2 A^2 = \frac{1}{2} m_{\rm a}
J_A/R^3$ and
the adiabatic invariant is just proportional to $R^3 \rho_{\rm a} /m_{\rm a}$,
the axion number per comoving volume.
We also see that the energy density in the adiabatic approximation scales
as $m_{\rm a} /R^3$, and so the evolving energy density remembers the initial
misalignment angle $\theta_{\rm in}=A(t<\tilde{t})$ via the invariant
$J_A=m_{\rm a}(\tilde{T})\theta_{\rm in}^2 f_{\rm a}^2 R(\tilde{T})^3$ in
the expression for the energy density:
\begin{equation}
\rho_{\rm a}(T)=\frac{1}{2} J_A m_{\rm a}(T)/R^3(T).
\label{eq:andens}
\end{equation}
The adiabatic approximation is invalid
near $\tilde{t}\,$ since $\dot{m}_{\rm a}$ cannot be $\ll m_{\rm a}^2$
at the onset of oscillations.
Furthermore, the adiabatic prediction of the axionic amplitude evolution
(time--independent $J_A \propto  R^3 \rho_{\rm a} /m_{\rm a}$)
implies that the axions behave like relativistic dust,
with no axions produced or annihilated.
This cannot be the case near or before $\tilde{t}\,$, not even for a
purely harmonic potential, since the overdamped solution is $\phi = constant$
in a \underline{physical volume},
which means that axions are constantly produced in the comoving volume!
On the other hand, the approximations are expected to rapidly become accurate
at times shortly beyond $\tilde{t}$.
To investigate the validity of the approximations made in the analytic
work of \cite{classical axion}, Turner \cite{Turner} integrated
Eq.\ (\ref{eq:eom}) numerically for a harmonic potential, as well as for the
cosine--shaped anharmonic axion potential. For initial conditions within the
harmonic part of the potential, he found a multiplicative factor $f_c(n)$
correcting the r.h.s.\ of Eq.\ (\ref{eq:andens}), which he fit with the linear
expression
\footnote{When one inputs the perturbative instanton result
$n\sim$ 3.6 to 3.8, one obtains a correction factor $f_c({\rm perturbative\,
instanton})\sim$ 1.34 to 1.39, surprisingly close to unity.  However,
the validity of the perturbative instanton approximation,
and therefore of the smallness of this correction factor, is suspect.}
\[
f_c(n)=0.44+0.25n\mbox{     (numerical fit).}
\]
(Recall that $n$ is the power characterizing the axion mass turn--on in Eq.\
(\ref{eq:m(T)})).
The correction factor is independent of $\theta_{\rm in}$ as a consequence
of the homogeneous property of Eq.\ (\ref{eq:eom}).
With our solution to Eq.\ (\ref{eq:eom}), we can in fact
derive the correction factor. Let us do so.
The solution for a
harmonic potential with a power--law time--dependence
(which is exactly our model potential
Eq.\ (\ref{eq:modelV}), as long as
$|\theta_{\rm in}|<\pi/2\,$), is
\[
\theta\left(t,|\theta_{\rm in}|<\pi/2\right)=
\theta_{\rm in}\Gamma\!\left(1+\frac{1}{2n+4}\right)\left(\frac{3}{2n+4}
\right)^{-\frac{1}{2n+4}}
t^{-\frac{1}{4}}J_{\frac{1}{2n+4}}
\left(\frac{3\,t^{\frac{n+2}{2}}}{n+2}\right)
\mbox\ .\]
The initial value $\theta(t\!=\!0)=\theta_{\rm in}$ determines the
normalization factor.
This result has, with different scaling of $t$, been published in ref.\
\cite{Dine}.
The proportionality constants are derived
from the series expansion of the Bessel functions
\begin{equation}
\left.
\begin{array}{r}
J_{\nu}(z)\\
I_{\nu}(z)
\end{array}\right\}=
\left(\frac{z}{2}\right)^\nu\sum_{n=0}^{\infty}\frac{(\mp1)^n}{n!\Gamma
(n+\nu+1)}\left(\frac{z}{2}\right)^{2n}\mbox\ .
\label{eq:seriesexpansion(z)}
\end{equation}
The asymptotic form of Bessel functions of the first kind
\begin{equation}
J_{\pm\nu}(z)\stackrel{z\rightarrow\infty}{\longrightarrow}
	\sqrt{\frac{2}{\pi z}}\,
\cos\left(z-(\pm 2\nu+1)\frac{\pi}{4}\right)\,+{\cal O}(z^{-3/2})
\label{eq:asymJ}
\end{equation}
translates the $J_{\frac{1}{2n+4}}$ solution
into the late time expression
\begin{eqnarray}
\theta(t)\stackrel{z\rightarrow\infty}{\longrightarrow}
\theta_{\rm in}\Gamma\!\left(1+\frac{1}{2n+4}\right)
\left(\frac{3}{2n+4}\right)^{-\frac{1}{2n+4}}
\sqrt{\frac{2n+4}{3\pi}}\,t^{-\frac{n+3}{4}}\nonumber\\[5mm]
\times\cos\left(\frac{3\,t^{\frac{n+2}{2}}}{n+2}-\frac{\pi}{4}
	(\frac{1}{n+2}+1)\right)
\mbox\ .\label{eq:convexasym}
\end{eqnarray}
The $n$--dependent factors in the amplitude provide the improvement of the
amplitude in Eq.\ (\ref{eq:ansatz}), and the square of these factors correct
the
adiabatic axionic mass density
presented on the r.h.s.\ of Eq.\ (\ref{eq:andens}), viz.
\begin{equation}
f_c=\Gamma^2\!\left(1+\frac{1}{2n+4}\right)
\left(\frac{3}{2n+4}\right)^{-\frac{1}{n+2}}\,
\left(\frac{2n+4}{3\pi}\right)\mbox\ .
\label{eq:f_c}
\end{equation}
Note that no approximations were made in deriving this non--adiabatic result.
There are two assumptions, namely, that the temperature dependence of the
axion mass is given by Eq.\ (\ref{eq:m(T)}), and that the effective potential
is purely quadratic (valid whenever the interactions, including
self--interactions, of the scalar fields are sufficiently weak).

The multiplicative correction factor $f_c$ is shown for $0\le n \le 10$
in Fig.\ 2.  The linear fit over the range of
these values is given by $f_c=0.456+0.247n$, in very good agreement with
the numerical results of \cite{Turner}. The near--linearity of the
correction factor in Eq.\ (\ref{eq:f_c}) is evident in our figure
over the entire range of $n$.

That $f_c$ depends only on $n$ and not on
$\theta_{\rm in}$ implies that
the probability distribution for $\theta$-angles in a harmonic potential
retains its shape for all times.
That is to say, the relative over-- and under--densities do not evolve.
If the initial misalignment angle $\theta=\overline{\theta}+\delta\theta$
is small enough to validate the harmonic approximation to the true
axion effective potential, then we may write
\begin{equation}
\frac{\delta\theta}{\overline{\theta}}=\frac{\delta\theta_{\rm
in}}{\overline{\theta
_{\rm in}}}\mbox\ ,\mbox\ {\rm for\:}\left|\overline{\theta_{\rm
in}}\right|<\frac{\pi}{2}\mbox\ .
\label{eq:deltheta}
\end{equation}
In particular, Gaussian amplitude fluctuations stay Gaussian after mass
turn-on.
This is not the case for fluctuations large enough to feel
the anharmonic effects.  ``Large enough" in our model potential
given by Eq.\ (\ref{eq:modelV}) means
$|\theta|>\frac{\pi}{2}$, so that the field's initial value tastes
the concave part of our potential.

\section{Large initial misalignment angles: anharmonic evolution}

If $|\theta_{\rm in}|>\frac{\pi}{2}$, we
fit the initial condition
\begin{equation}
\epsilon_0(t_{\rm o}\!\ll 1)=\epsilon_{\rm in}
\label{eq:DELTA}
\end{equation}
to the solution in (\ref{eq:solution.b}) with positive index,
since the negative index solution is singular at $t=0$.
It may be that the initial value for $\dot{\epsilon}(t_{\rm o})$
is also a stochastic parameter, smeared by the
equal--time canonical commutator $[\theta,\dot{\theta}]=i$ of the
quantized theory.  However,
rather than complicate this analysis, we will assume that
$\dot{\epsilon}(t_{\rm o})$ is correctly given by omitting the negative order
Bessel function at $t_{\rm o}$.
At the time $t_1$ when $|\epsilon_0(t_1)|=\frac{\pi}{2}$,
$\epsilon_0$ must be matched to $\theta_1(t)$,
which is a linear combination of the two functions
in (\ref{eq:solution.a}); also, the first
derivatives must be matched in the fashion
$\dot{\epsilon}_0(t_1)=-\dot{\theta}_1(t_1)$.
For an initial condition not close to $|\theta_{\rm in}|=\pi$,
matching at the field value $\frac{\pi}{2}$ will only be necessary once,
since the oscillation begins while the
Hubble term is significant, acting as friction to reduce the field amplitude
(as $R^{-3/2} \propto t^{-3/4}$ in the adiabatic approximation).
The further evolution of the axion field then happens in the convex part of the
potential, and is given by the functions in (\ref{eq:solution.a}).
We will see that it suffices for quite a wide range of initial conditions to
only calculate $\theta_1(t)$.

If, however, $|\theta_{\rm in}|$ is close to $\pi$, then
the solution $\epsilon_0$ is metastable and so begins a late oscillation when
the Hubble term is smaller; the matched solution
$\theta_1$ eventually evolves beyond the convex part of the
effective potential.  This necessitates a second matching, of
$\theta_1(t_2) = -\frac{\pi}{2}$ to $\epsilon_2(t_2)$ and
$\dot{\theta}_1(t_2)$ to $-\dot{\epsilon}_2(t_2)$, where $\epsilon_2(t)$
is a linear combination of functions in (\ref{eq:solution.b}),
and a third matching of $\epsilon_2(t_3) = -\frac{\pi}{2}$ to
$\theta_3(t_3)$ and $\dot{\epsilon}_2(t_3)$ to $-\dot{\theta}_3(t_3)$,
where $\theta_3(t)$ is a new linear combination of functions in
(\ref{eq:solution.a}).
A schematic illustration of an oscillation pattern is given in Fig.\ 1.
For large enough $|\theta_{\rm in}|$, the harmonic minimum of the effective
potential will be overshot arbitrarily often.  This is because
an initial field amplitude chosen near the
maximum of the potential will outlive the demise of the damping
term in Eq.\ (\ref{eq:geneq}) which decreases rapidly with time.
Said another way,
a classical field amplitude of exactly $\pi$ will remain at $\pi$ and
not oscillate at all!  And so an initial amplitude slightly removed from
$\pi$ will begin late oscillations when the damping/expansion is insignificant,
and then oscillate for almost ever, in accord with energy conservation.

For the construction of the individual solutions we use the notational
convention
\begin{equation}
\begin{array}{lll}
\epsilon_0(t)=\pi-\theta(t)\;>0 & \mbox{for } t_{\rm o}<t<t_1\mbox\ ;&  \\[1mm]
\theta_1(t)=\theta(t) & \mbox{for } t_1<t<t_2\mbox\ ,& \mbox{ with }
\theta_1(t_1)=\epsilon_0(t_1)=\frac{\pi}{2}\mbox\ ;\\[1mm]
\epsilon_2(t)=-\pi-\theta(t)<0 & \mbox{for } t_2<t<t_3\mbox\ ,& \mbox{ with }
\epsilon_2(t_2)=\theta_1(t_2)=-\frac{\pi}{2}\mbox\ ;\\[1mm]
\theta_3(t)=\theta(t) & \mbox{for } t_3<t<t_4\mbox\ ,& \mbox{ with }
\theta_3(t_3)=\epsilon_2(t_3)=-\frac{\pi}{2}\mbox\ ;\\[1mm]
\epsilon_4(t)=\pi-\theta(t)>0 & \mbox{for } t_4<t<t_5\mbox\ ,& \mbox{ with }
\epsilon_4(t_4)=\theta_3(t_4)=\frac{\pi}{2}\mbox\ ,\\[1mm]
\vdots & \vdots & \vdots \\
\end{array}
\label{eq:X}
\end{equation}
which is illustrated on the scale in Fig.\ 1.
The series terminates at
$\theta_N(t)=\theta(t)$
where $\theta_N(t)=(-1)^{\frac{N+1}{2}}\frac{\pi}{2}$
has no solution for $t>t_{N}$.
So the subscripts on $\epsilon$ and $\theta$ display a running count of
the number of times $t_m$ that matching of solutions at the values
$\theta=\pm \pi/2$ is required; equivalently, the index on $\theta_m$ minus
one is twice the number of times
that the oscillating field has risen above the convex potential into the
anharmonic regime.

It is useful to introduce a new time scaling
\begin{equation}
z\equiv\frac{3\,t^{\frac{n+2}{2}}}{n+2}.
\label{eq:BOX}
\end{equation}
In terms of $z$, the general forms of the solutions in eqns. (7) and (8) become
\begin{equation}
\begin{array}{rcl}
\theta_m (z) & = & z^{-\nu}\left[ A_m J_{\nu}(z) + B_m
J_{-\nu}(z)\right]\\[2mm]
\epsilon_n(z) & = & z^{-\nu}\left[C_n I_{\nu}(z)+D_n I_{-\nu}(z)\right]\mbox\
,\\
\end{array}
\label{eq:CIRCLE}
\end{equation}
where $\nu=(2n+4)^{-1}$.
Physical time $t$ is obtained from $z$--time by inverting Eq.\ (\ref{eq:BOX}):
\begin{equation}
t=\left(\frac{(n+2)z}{3}\right)^{\frac{2}{n+2}}.
\label{eq:TIME}
\end{equation}
{}From Eq.\ (\ref{eq:X}) and the continuity of $\dot{\theta}$
it is clear that the connectivity
conditions for the time derivatives are simply
\footnote{Since the common prefactor $z^{-\nu}$ in Eq.\ (\ref{eq:CIRCLE})
is a $C^{\infty}$
function at all $z>0$, we may replace the $\epsilon_m$ and $\theta_m$
in the matching conditions with the simpler and equivalent
$z^{\nu}\epsilon_n = (C_n I_{\nu}(z)+D_n I_{-\nu}(z))$
and $z^{\nu}\theta_m = (A_m J_{\nu}(z) + B_m J_{-\nu}(z))$.
However, the recurrence relations on this page make it equally simple to work
with the original functions.}
(with $m$ an odd integer here),
\begin{equation}
\dot{\epsilon}_{m-1}(z_m)=-\dot{\theta}_m(z_m),\;{\rm and}\;
\dot{\theta}_m(z_{m+1})=-\dot{\epsilon}_{m+1}(z_{m+1}).
\label{eq:PLUS}
\end{equation}
To compute these derivatives, some recurrence relations are useful.
For $Z_{\nu}$ any Bessel function, we have
\[
\frac{d}{dz}\left(z^{-\nu}Z_{-\nu}(z)\right)=z^{-\nu}Z_{-(\nu+1)}(z)\mbox\ ,
\]
while
\[
\frac{d}{dz}\left(z^{-\nu}J_{\nu}(z)\right)=-z^{-\nu}J_{\nu+1}(z)\mbox\ ,
\]
and
\[
\frac{d}{dz}\left(z^{-\nu}I_{\nu}(z)\right)=z^{-\nu}I_{\nu+1}(z)\mbox\ .
\]

To compute the coefficients $A_m$, $B_m$, $C_n$, $D_n$, ($m$ odd, $n$ even),
the
procedure is now clear. The initial condition $\epsilon_0(0)=\epsilon_{\rm in}$
fixes the normalization of $\epsilon_0(z)$ to be
\[
\epsilon_0(z) = \epsilon_{\rm in}\Gamma(1+\nu)\left(\frac{2}{z}\right)
	^{\nu}I_{\nu}(z)\mbox\ ;
\]
that is, $C_0=\epsilon_{\rm in}\Gamma(1+\nu)2^{\nu}$ and $D_0=0$.
Next, $z_1=\frac{3}{n+2}t_1^{\frac{n+2}{2}}$ is found by solving
$\left.\epsilon_0(z)\right|_{z=z_1}=\frac{\pi}{2}$. The conditions $\theta_1
(z_1)=\frac{\pi}{2}$ and $\dot{\epsilon}_0(z_1)=-\dot{\theta}_1(z_1)$ translate
in terms of the notation in Eq.\ (\ref{eq:CIRCLE}) into
\begin{equation}
\begin{array}{rcl}
A_1 J_{\nu}(z_1) + B_1 J_{-\nu}(z_1) & = & \frac{\pi}{2}z_1^{\nu}\\[2mm]
A_1 J_{\nu+1}(z_1) - B_1 J_{-\nu-1}(z_1) & = & C_0 I_{\nu+1}(z_1)
\end{array}
\label{eq:MATCH}
\end{equation}
The inverse of
\[
M= \left( \begin{array}{cc}
	J_{\nu}(z) & J_{-\nu}(z) \\
	J_{\nu+1}(z) & -J_{-(\nu+1)}(z) \\
\end{array} \right)
\]
is
\[
M^{-1}= \frac{1}{W(J_{\nu},J_{-\nu};z)}
\left( \begin{array}{cc}
	J_{-(\nu+1)}(z) & J_{-\nu}(z) \\
	J_{\nu+1}(z) & -J_{\nu}(z) \\
\end{array} \right),
\]
where $W(J_{\nu},J_{-\nu};z)=
J_{\nu}(z)J_{-(\nu+1)}(z)+J_{-\nu}(z)J_{\nu+1}(z)=-\frac{2\sin\pi\nu}{\pi z}$
is the Wronskian.
Thus,
\begin{equation}
\left(\begin{array}{c}A_1\\ B_1 \end{array}\right) =
	-\frac{\pi z_1}{2\sin \pi \nu}
\left( \begin{array}{cc}J_{-(\nu+1)}(z_1) & J_{-\nu}(z_1)\\
	J_{\nu+1}(z_1) & -J_{\nu}(z_1) \end{array}\right)
\left( \begin{array}{c}
	\frac{\pi}{2}z_1^{\nu}\\
	C_0 I_{\nu+1}(z_1)\\
\end{array}\right)\mbox\ ,
\label{eq:00}
\end{equation}
and $\theta_1(z)=A_1 J_{\nu}(z) + B_1 J_{-\nu}(z)$.
Then we test whether or not the equation $\theta_1(z)=-\frac{\pi}{2}$ has
solutions $z>z_1$\ . If it does not, then $\theta_1(z)$ is the final solution
for the field amplitude.
If there are further solutions to the transcendental equation in $z$,
then we name the smallest of these solutions $z_2$,
and derive
\begin{equation}
\begin{array}{l}
\left(\begin{array}{c}C_2\\ D_2 \end{array}\right) =
	{\displaystyle -\frac{\pi z_2}{2\sin \pi \nu}\times} \\[6mm]
\left( \begin{array}{cc}-I_{-(\nu+1)}(z_2) & -I_{-\nu}(z_2)\\
	I_{\nu+1}(z_2) & I_{\nu}(z_2) \end{array}\right)
\left( \begin{array}{c}
	\frac{\pi}{2}z_2^{\nu}\\
	A_1 J_{\nu+1}(z_2)-B_1 J_{-(\nu+1)}(z_2)\\
\end{array}\right)\mbox\ ,
\label{eq:ELLIPSE}
\end{array}
\end{equation}
using the Wronskian formula
\[
W(I_{\nu},I_{-\nu};z)=
	I_{\nu}(z)I_{-(\nu+1)}(z)-I_{-\nu}(z)I_{\nu+1}(z)=
	-\frac{2\sin\pi\nu}{\pi z}\mbox\ .
\]
One can repeat this procedure as many times as needed
to calculate all the further coefficients.
For example, $A_3, B_3, C_4,$ and $D_4$ are given by equations identical to
(\ref{eq:00}) and (\ref{eq:ELLIPSE}) after some straightforward substitutions:
\\
(i) each subscript on $A, B, C, D,$ and $z$ is incremented by two, \\
(ii) $C_0 I_{\nu+1}$ in the column vector on the right--hand side of
(\ref{eq:00}) is changed to $C_2 I_{\nu+1} + D_2 I_{-(\nu+1)}$, and \\
(iii) the matching value of the field, $\frac{\pi}{2}$
in the column vectors on the right--hand
sides of eqs. (\ref{eq:00}) and (\ref{eq:ELLIPSE}), is changed to the new
matching value, $-\frac{\pi}{2}$. \\

The final field amplitude
$\theta_N(\epsilon_{\rm in},t)=z^{-\nu}[A_N J_{\nu}+B_N J_{-\nu}]$
is generated
\footnote{Prior calculations used step-by-step integration of the field
equation.  When multi--oscillations are present, such an approach is
computationally intensive.  We proceed with a simple root--finder algorithm.
We find qualitative agreement \cite{thesis} with the earlier work
\cite{Turner}.} in this recursive way.
We show $z_1(\epsilon_{\rm in})$ in Fig.\ 4.  It is seen that the values of
$z_1$ hardly depend on $n$, whereas $t_1$, which is an $n$-dependent
reparameterization of $z$, given in Eq.\ (\ref{eq:TIME}), depends sensitively
on $n$.  We also show the occurences and values of $z_2, z_3, z_4$ and $z_5$
as a function of $\epsilon_{\rm in}$, for values of $n = 0, 3.7, 8$.
Several important inferences may be drawn from Fig.\ 4:\\
(i) the convex region is overshot (for $n=3.7$)
when $\epsilon_{\rm in}\stackrel{<}{\sim}0.02\,\pi$, and double overshooting of
the
convex region occurs when $\epsilon_{\rm in}\stackrel{<}{\sim}10^{-3}\,\pi$;\\
(ii) the $z$--time spent in the overshot concave region,
$z_3-z_2$, is relatively short;\\
(iii) after a $z$--time equal to the largest $z_i$ in the figure, the field
oscillates in the harmonic region and no further growth of density
perturbations
takes place; the largest $z_i$ is relatively small on the cosmic time scale;\\
(iv) we do not need to cut off the power--law growth of mass in our
calculations, since the anharmonic effects which govern density
growth terminate before the axion mass reaches its
low--temperature value:
with the power--law growth of Eq.\ (\ref{eq:m(T)}),
the mass saturates at a $z$ value of $z_c=(\frac{3}{n+2})t_c^{\frac{n+2}{2}}=
(\frac{3}{n+2})(m_{\rm a}(\infty)/m_{\rm a}(\tilde{t}))^{\frac{n+2}{n}}$.
The exponent ${\frac{n+2}{n}}$ is always $>1$. Thus, we expect that
$z_c > (\frac{3}{n+2})(m_{\rm a}(\infty)/m_{\rm a}(\tilde{t}))$.
Since $m_{\rm a}(\infty)$ is $\geq 10^{-5}$ eV, and $m_{\rm a}(\tilde{t})\simeq
3H(T\simeq 1 \rm{GeV})\simeq 10^{-9}$ eV, we have
$z_c \stackrel{>}{\sim} 10^{4}$, much larger than the occurrences of
the matching times $z_1$ through $z_5$, even for infinitesimal values
of $\epsilon_{\rm in}$.
\footnote
{Once the mass has saturated at its zero--temperature value,
the growth index $n$ must go to zero, and the index and arguments
of the Bessel solutions become fixed at 1/4 and $\frac{3}{2}t$, respectively.
However, since the field evolution has entered
its harmonic epoch prior to the onset of these solutions,
these solutions are irrelevant to the growth of density perturbations.
Moreover, since constant mass implies adiabaticity, there is no further
increase in axion number.}
In practice, the basis \{$I_{\nu}, K_{\nu}$\} offers an advantage over
the basis  \{$I_{\pm\nu}$\} for expanding the $\epsilon(z)$ solutions.
This is because the asymptotic expressions
for the former differ exponentially
while those of the latter are in fact identical (being
functions of $\nu^2$ rather than of $\nu$):
\begin{equation}
I_{\pm\nu}(z)\stackrel{z\rightarrow\infty}{\longrightarrow}
	\frac{e^z}{\sqrt{2\pi z}}\,
	\left(1-\frac{4\nu^2-1}{8z}+{\cal O}(z^{-2})\right)
\label{eq:asymI}
\end{equation}
and
\begin{equation}
K_{\nu}(z)\stackrel{z\rightarrow\infty}{\longrightarrow}
	\sqrt{\frac{\pi}{2 z}}\,e^{-z}\,
	\left(1+\frac{4\nu^2-1}{8z}+{\cal O}(z^{-2})\right)
\label{eq:asymK}
\end{equation}
The use of $K_{\nu}$ rather than $I_{-\nu}$ gives much better stability
for numerical work, and we have in fact used it.
The substitution of $K_{\nu}$ for $I_{-\nu}$ in the equations above
for the $A, B, C, D$ coefficients is effected with the following three changes:
\\
(i) $I_{-\nu}$ is replaced with $K_{\nu}$; \\
(ii) $I_{-(\nu+1)}$ is replaced with $-K_{\nu+1}$,
the relative minus sign arising from the fact that $K_{\nu}$ obeys the same
derivative--recurrence relation as $J_{\nu}$,
opposite in sign to that of $I_{-\nu}$; \\
(iii) the Wronskian $W(I_{\nu},I_{-\nu};z)=-\frac{2\sin\pi\nu}{\pi z}$
is replaced with the Wronskian
$W(I_{\nu},K_{\nu};z)=
        I_{\nu}(z)K_{\nu+1)}(z)+K_{\nu}(z)I_{\nu+1}(z)=1/z.$

The final values $A_N$ and $B_N$ are sufficient to determine the final
density contrast, as we now show.
{}From the asymptotic expansion of $J_{\nu}$ we can compute the
amplitude $\theta_{\rm out}$ at large $z$:
\begin{equation}
\begin{array}{ll}
\theta_N (z) =&z^{-\nu}\left[ A_N J_{\nu}(z) + B_N J_{-\nu}(z)\right]\\
	&\stackrel{z\rightarrow\infty}{\longrightarrow}
		z^{-\nu}\sqrt{\frac{2}{\pi z}}
\left[ A_N \cos(z-\frac{\pi}{4}(1+2\nu)) + B_N  \cos(z-\frac{\pi}{4}(1-2\nu))
\right]\\
	&=\theta_{\rm out}\cos(z-\delta),
\end{array}
\end{equation}
with
\begin{equation}
\begin{array}{ll}
\theta^2_{\rm out}=&\left(\frac{2}{\pi z^{2\nu+1}}\right)
\left[(A_N+B_N)^2 \cos^2\left(\frac{\nu\pi}{2}\right)+
    (A_N-B_N)^2 \sin^2\left(\frac{\nu\pi}{2}\right)\right]\\
&=\frac{2}{\pi}\left(\frac{n+2}{3}\right)^{\frac{n+3}{n+2}} t^{-\frac{n+3}{2}}
\left[A_N^2 + B_N^2 + 2A_N B_N\cos(\frac{\pi}{2n+4})\right]
\end{array}
\label{eq:FINAL}
\end{equation}
and
\[
\delta=\frac{\pi}{4}+\arctan\left[(\frac{A_N-B_N}{A_N+B_N})
	\tan(\frac{\nu\pi}{2})
\right].
\]
Note that the final coefficients $A_N$ and $B_N$ carry a memory of
all the matching conditions, i.e. $A_N=A_N(z_1, z_2, ...,z_N)$
and similarly for $B_N$;
in turn, the $z_i$ are predetermined by the value of the exponent $n$, and by
the initial amplitude $\theta_{\rm in}=\pm\pi-\epsilon_{\rm in}$.
Finally, we may write down the asymptotic energy density appropriate when the
field is evolved into the harmonic region of the potential.
It is just $\rho_{\rm a}=\frac{1}{2}f_{\rm a}^2 m_{\rm a}^2(t)\theta_{\rm
out}^2$,
where $\theta_{\rm out}^2$ is the squared field amplitude of
Eq.\ (\ref{eq:FINAL}),
and $m_{\rm a}^2(t)=m_{\rm a}^2(1)t^n$.

It is interesting to compare our final solution in Eq.\ (\ref{eq:FINAL}) to
the results of the harmonic approximation, with a time--independent axion mass,
and with an adiabatically growing mass.
The fixed mass solution to the harmonic Eq.\ (\ref{eq:eom})
is simply
$\theta_{\rm out}=\theta_{\rm in} \Gamma(\frac{5}{4})
	(\frac{3t}{4})^{-\frac{1}{4}}J_{\frac{1}{4}}(m_{\rm a} t)$.
The approximate answer is simple, but essential
physics (mass turn--on, and anharmonicity) has been left out.
The adiabatic solution to the harmonic potential,
presented in Eq.\ (\ref{eq:ansatz})
and explained thereafter, yields
$\theta^2 \propto (m_{\rm a} R^3)^{-1}$, and so
$\theta_{\rm out}^2=\theta_{\rm in}^2 \,(\tilde{t}/t)^{\frac{n+3}{2}}$.
The asymptotic time--dependence of course agrees with our more careful answer,
since in any model the field will evolve into the harmonic minimum of the
potential at late times.  However, the overall magnitude is different,
because the harmonic approximation ignores important physics, at least for
large $\theta_{\rm in}$.  Finally, we comment that the solution to
Eq.\ (\ref{eq:geneq}) when the expansion term is omitted is
$\sqrt{t}\, Z_{\pm\frac{1}{n+2}}(\frac{3}{n+2}t^{\frac{n+2}{2}})$, with
$Z$ any Bessel (modified Bessel) function for the convex (concave) potential.
The time dependence of this solution, and even the asymptotic behavior of
the oscillation amplitude, $\sim t^{-n/4}$, show how much damping due to
expansion affects our true solution.

The correction factor $f_c(\theta_{\rm in})$ for the axion
energy density from the adiabatic, harmonic solution is just the ratio of
$\theta_{\rm out}^2$ in Eq.\ (\ref{eq:FINAL}) to the harmonic result,
\begin{equation}
f_c(\theta_{\rm in})=\frac{2}{\pi}\left(\frac{n+2}{3}\right)^{\frac{n+3}{n+2}}
\left[A_N^2 + B_N^2 + 2A_N B_N\cos(\frac{\pi}{2n+4})\right]
\label{eq:fc}
\end{equation}
This correction factor is plotted in Fig.\ 3, for three values of $n$.
The correction exceeds 2 when $\epsilon_{\rm in}/\pi \stackrel{<}{\sim} 0.07$,
and exceeds 5 when $\epsilon_{\rm in}/\pi \stackrel{<}{\sim} 3 \times 10^{-3}$.

To quantify the final axion density contrast,
$\frac{\delta{\rho_{\rm a}}}{\overline{\rho_{\rm a}}}$,
we find it convenient to define its dependence on
the initial field misalignment by
\begin{equation}
\frac{\delta \rho_{\rm a}}{\overline{\rho_{\rm a}}}=
\frac{1}{2}\xi(\overline{\theta_{\rm in}})
\frac{\delta\theta_{\rm in}^2}{\overline{\theta_{\rm in}^2}}
\approx \xi(\overline{\theta_{\rm in}})
\frac{\delta\theta_{\rm in}}{\overline{\theta_{\rm in}}}
\label{eq:xi}
\end{equation}
The density fluctuation $\delta\rho$ about the mean density
$\overline{\rho}$ is defined by $\rho=\overline{\rho}+\delta\rho$.
The approximate expression in Eq.\ (\ref{eq:xi}) is valid for small
fluctuations.
In the harmonic regime ($\theta_{\rm in}<\frac{\pi}{2}$ in our model),
$\rho_{\rm a} \propto \theta^2_{\rm in}$,
and so $\xi=2$. For $|\theta_{\rm in}|>\frac{\pi}{2}$,
we expect anharmonic effects to retard the
progression of the amplitude toward the minimum of the potential,
thereby effecting a larger density contrast, $\xi(\theta_{\rm in})>2$.
In the extreme, we expect
$\delta \rho_{\rm a}$ in our analysis to diverge as
$|\theta_{\rm in}|\rightarrow \pi$,
because at the maximum of the potential there is no force to push or pull
the field toward its minimum.
However, in reality there are other limitations to this formalism
at the potential maximum.  For example, a field at a potential maximum
will in general evolve in all directions characterized by a negative gradient.
This then leads to domain walls.
The energetics and evolution of domain walls are dominated
by field gradient terms, and so are outside of our zero--mode
formulation.  As mentioned previously, the existence of walls
conflicts with the deductions of observational cosmology.

Since $\frac{\delta\rho}{\overline{\rho}}$ is just
$\frac{\delta\theta_{\rm out}^2}{\overline{\theta_{\rm out}^2}}$,
we have $\xi(\overline{\theta_{\rm in}})=2\frac{\overline{\theta_{\rm in}^2}}
{\overline{\theta_{\rm out}^2}}\frac{\delta\theta_{\rm
out}^2}{\delta\theta_{\rm in}^2}$.
For small fluctuations,
\begin{equation}
\xi(\overline{\theta_{\rm in}}) \approx 2\frac{\overline{\theta_{\rm in}}}
{\overline{\theta_{\rm out}}}\frac{d\theta_{\rm out}}{d\theta_{\rm in}} =
2\frac{\overline{\theta_{\rm in}}}{\overline{\epsilon_{\rm in}}}
\left(-\frac{d\;\ln(\theta_{\rm out})}{d\;\ln(\epsilon_{\rm in})}\right).
\label{eq:deriv}
\end{equation}
(Recall that $\overline{\epsilon_{\rm in}}=\pm\pi-\overline{\theta_{\rm in}}$.)
Using our mostly analytic procedure
we have computed values of $\xi$ for cases where
the convex part of the effective potential is overshot once.
We do so by computing
$\theta_{\rm out}$ as a function of $\epsilon_{\rm in}$ within our model,
and then taking the derivative numerically to determine $\xi$ as given
in Eq.\ (\ref{eq:deriv}).

The advantages of our method over
straightforward numerical integration
\cite{Turner,thesis}, are
numerical stability and speed.
Speed and stability are especially advantageous for the most interesting
case of a large initial misalignment angle.
We use numerics only to find roots of a few transcendental equations and to
perform 2 by 2 matrix multiplication to obtain the
Bessel coefficients in Eqs.\ (\ref{eq:00}), (\ref{eq:ELLIPSE}),
and their successors.
After we obtain the values of the final coefficients $A_N$ and $B_N$,
$\theta_{\rm out}^2$ is given by Eq.\ (\ref{eq:FINAL}).
We show this final enhancement of the density--fluctuations in Figs.\ 5 and 6.
In Fig.\ 5 the divergence arising from the factor
$\overline{\epsilon_{\rm in}}^{-1}$
in Eq.\ (\ref{eq:deriv}) is scaled away and our
results are compared with previous
ones \cite{Lyth}, and with our asymptotic solution, Eq.\ (\ref{eq:xi2}).
In Fig.\ 6 we show the exact solution of our model for different values of $n$.
The enhancement, $\xi/2$, is large for small $\epsilon_{\rm in}$;
e.g. it is a factor
of $\sim$ (2,3,4,13) for $\epsilon_{\rm in}\sim
(0.15,0.10,0.05,0.01)\times\pi$.
This is our most important result.

If one were to abandon the small fluctuation condition
$\delta\theta\ll\overline{\theta}$, then one might seek to compare
the enhancements of densities arising from very different initial
misalignment angles.  The relative density enhancement is obtained by
integrating Eq.\ (\ref{eq:xi}).  The result is
\[
\frac{\rho_2}{\rho_1}=\exp \left[\int_{\theta_{{\rm in},1}}^{\theta_{{\rm
in},2}}
\xi(\theta)\frac{d\,\theta}{\theta}\right].
\]
{}From Fig.\ 6, it is evident that such an integral can become very large.
Since $\xi(\theta)$ grows rapidly with increasing large $\theta$,
we may write down an approximate result:
$\frac{\rho_2}{\rho_1}\approx \exp[\xi(\theta_2)\ln(\theta_2/\theta_1)]
=\left(\frac{\theta_2}{\theta_1}\right)^{\xi(\theta_2)}$.
However, large contrasts resulting from large initial field misalignments
may not be trustworthy without inclusion of the gradient term \cite{KT}.

We will return to further
discussion of the results displayed in Fig.\ 5 in the next sections.

\section{An asymptotic approximation}
We have convinced ourselves that for a large range of
$\epsilon_{\rm in}$ values
we get a reasonable approximation for the density
contrast by including only the first occurence of $\theta(z)$, i.e.\
$\theta_1(z)$, and by using the asymptotic expression
for the Bessel solutions $\epsilon_0(z)$ and $\theta_1(z)$
in the initial concave and the final convex regions, respectively.
This may be so because for small $\epsilon_{\rm in}$,
$z_1$ is $\gg 1$, allowing the curvature of the potential to grow to
a large value before the onset of oscillations.
This in turn implies a high oscillation frequency, so that
the true solution simply does not
spend much time in the concave potential (the duration of the
$\epsilon_2$ epoch is $z_3-z_2$)
after it overshoots the convex potential.
These interpretations are supported by the $z_1,\,z_2,\mbox{ and } z_3$ values
in Fig.\ 4.
Let us then pursue this approximate scheme of
using only asymptotic $\epsilon_0$ and asymptotic $\theta_1$.

Asymptotic $\epsilon_0$ is given by
\begin{equation}
\epsilon_0(z)=C_0 z^{-\nu} I_{\nu}(z)\stackrel{z\gg 1}{\longrightarrow}
C_0 z^{-\nu}\frac{1}{\sqrt{2\pi z}}\mbox{e}^z ,
\label{eq:concaveasym}
\end{equation}
with $C_0=\epsilon_{\rm in}2^{\nu} \Gamma(\nu+1)$.
This equation is equivalent to the functional form
obtained by Lyth \cite{Lyth}, which he heuristically
based on inverting the convex potential to a concave potential by
replacing time with imaginary time, and
applying the adiabatic approximation; equivalently,
the asymptotic cosine solution is replaced by an asymptotic $\cosh$.
\footnote{The multiplicative correction factor to the adiabatically--evolving
density fluctuation in the harmonic concave potential
is exactly the same as the one obtained in Eq.\ (\ref{eq:f_c})
for the harmonic convex regime, which we found to be of order one.}
Thus, the satisfactory behavior of our asymptotic solution supports the
approximations made in \cite{Lyth}.
As long as the field remains in the concave regime, the field equation is
linear and the density contrast remains constant:
\[
\frac{\delta\epsilon}{\overline{\epsilon}}=\frac{\delta\epsilon_{\rm in}
}{\overline{\epsilon_{\rm in}}}\mbox\ .
\]
{}From setting the asymptotic solution (\ref{eq:concaveasym}) equal to
$\frac{\pi}{2}$, we get the equation whose solution is
the matching time $z_1$:
\begin{equation}
\frac{C_0}{\sqrt{2\pi}}z_1^{-(\nu+\frac{1}{2})}\mbox{e}^{z_1}
=\frac{\pi}{2}\mbox .
\label{eq:z_0}
\end{equation}
Even for this simple asymptotic solution
(and even when the convex potential is not overshot!),
the late--time density fluctuations cannot be
expressed as an analytic function of $\epsilon_{\rm in}$,
because the matching time $z_1(\epsilon_{\rm in})$ is a solution to the
transcendental equation (\ref{eq:z_0}).
Values for $z_1$ obtained with these asymptotic formulae
are indistinguishable from the solution to the
exact equation
$\epsilon_{\rm in} \Gamma(1+\nu)(\frac{2}{z_1})^{\nu} I_{\nu}(z_1)
=\frac{\pi}{2}$
shown in Fig.\ 4; from Eq.\ (\ref{eq:asymI}), the relative difference is
$(1-4\nu^2)/8z_1$.

We may write the asymptotic
solution in the convex regime as $z^{-\nu}$ times a linear combination of the
asymptotic forms of $J_{\pm \nu}$ given
in Eq.\ (\ref{eq:asymJ}), or alternatively, as
\begin{equation}
\theta(z)\stackrel{z\gg 1}{\longrightarrow}
A(\epsilon_{\rm in},n)\sqrt{\frac{2}{\pi}}z^{-(\nu+\frac{1}{2})}\cos(z-\delta)
\label{eq:theta}
\end{equation}
with
$\delta=
	\frac{\pi}{4}+\arctan(\frac{A_1-B_1}{A_1+B_1}\tan(\frac{\nu\pi}{2}))$.
The asymptotic phase $\delta$ is irrelevant for the calculation of the
mass-density spectrum.
{}From matching the field value in Eq.\ (\ref{eq:theta})
to $\pi/2$ at $z=z_1$, we get
\begin{equation}
A(\epsilon_{\rm in},n)\sqrt{\frac{2}{\pi}}\cos(z_1-\delta)=
\frac{\pi}{2}z_1^{\nu+\frac{1}{2}}\mbox\ .
\label{eq:match}
\end{equation}
To match the field derivatives at $z_1$ we need
\begin{equation}
\frac{d}{dz}\theta(z) \stackrel{z\gg 1}{\longrightarrow}  -A(\epsilon_{\rm
in},n)
\sqrt{\frac{2}{\pi}}z^{-(\nu+\frac{1}{2})}\left(\frac{\nu+\frac{1}{2}}{z}
\cos(z-\delta)+\sin(z-\delta)\right)\mbox\ ,
\label{eq:thetadot}
\end{equation}
and from Eq.\ (\ref{eq:concaveasym}),
\begin{equation}
\frac{d}{dz}\epsilon(z) \stackrel{z\gg 1}{\longrightarrow}
\left(1-\frac{\nu+\frac{1}{2}}{z}\right) \epsilon(z)
\mbox\ .
\label{eq:edot}
\end{equation}
{}From Eq.\ (\ref{eq:match})
and the matching of the derivatives, Eqs.\ (\ref{eq:thetadot})
and (\ref{eq:edot}) with a relative
minus sign, we eliminate the oscillating terms and
obtain the time independent coefficient of the
asymptotic amplitude $\theta$:
\[
A(\epsilon_{\rm in},n)\sqrt{\frac{2}{\pi}}
=\frac{\pi}{2}z_1^{\nu+\frac{1}{2}}\sqrt{2-\frac{4\nu+2}
{z_1}+\frac{(2\nu+1)^2}{z_1^2}} \mbox\ .
\]
The $n$--dependence of $A$ is implicit in $\nu(n)$ and the
$\epsilon_{\rm in}$--dependence is implicit in $z_1(\epsilon_{\rm in})$.
The axionic energy density
$\rho_{\rm a}=\frac{1}{2}f_{\rm a}^2(\dot{\theta}^2+m_{\rm a}^2\theta^2)$
is then equal to
$z_1^{2\nu+1}\left(2-\frac{4\nu+2}{z_1}+\frac{(2\nu+1)^2}{z_1^2}\right)$,
times an irrelevant time--dependent factor deducible from Eq.\
(\ref{eq:theta}).
Therefore,
\[
\frac{\delta\rho_{\rm a}(t)}{\overline{\rho_{\rm a}(t)}}=(2\nu+1)
\left[1+\frac{\frac{2}{\overline{z_1}}-\frac{2(2\nu+1)}{\overline{z_1}^2}}
{2-\frac{4\nu+2}{\overline{z_1}}+\frac{(2\nu+1)^2}{\overline{z_1}^2}}\right]\frac{\delta z_1}
{\overline{z_1}}\simeq(2\nu+1)(1+\frac{1}{\overline{z_1}})\frac{\delta z_1}
{\overline{z_1}}\mbox\
\]
for $\overline{z_1} \gg 1$.

With the relation
\begin{equation}
\frac{\delta\epsilon_{\rm in}}{\overline{\epsilon_{\rm in}}}=-\frac{\delta z_1}
{\overline{z_1}}\left[\overline{z_1}-(\nu+\frac{1}{2})
\right]\mbox\ ,
\label{eq:delepsilon}
\end{equation}
derived from Eq.\ (\ref{eq:z_0}),
the relative density fluctuations in terms of the initial misalignment
fluctuations are then given by
\begin{equation}
\frac{\delta \rho_{\rm a}(t)}{\overline{\rho_{\rm a}(t)}}\simeq
\frac{-(2\nu+1)}{\overline{z_1}-(\nu+\frac{1}{2})}
\left[1+\frac{1}{\overline{z_1}}\right]
\frac{\delta\epsilon_{\rm in}}{\overline{\epsilon_{\rm in}}}\mbox\ .
\label{eq:xi2}
\end{equation}
Notice that this means that for very small $\overline{\epsilon_{\rm in}}$
(implying large $z_1$)
\[
\frac{\delta \rho_{\rm a}(t)}{\overline{\rho_{\rm a}(t)}}\simeq
\frac{-(2\nu+1)}{\overline{z_1}}
\frac{\delta\epsilon_{\rm in}}{\overline{\epsilon_{\rm in}}},
\]
or in the notation of eq. (\ref{eq:xi}),
\[
\xi(\overline{\theta_{\rm in}})=\frac{2\nu+1}{\overline{z_1}}
\frac{\overline{\theta_{\rm in}}}{\pi-\overline{\theta_{\rm in}}}\mbox\ .
\]
We see then, that for $\theta_{\rm in}$ near $\pi$, amplitude fluctuations
evolve
to give density fluctuations enhanced over the harmonic result by the factor
\begin{equation}
\frac{1}{2}\xi(\overline{\theta_{\rm in}}) \simeq
(\frac{\pi}{\pi-\overline{\theta_{\rm in}}})(\frac{1}{2\overline{z_1}})\mbox\ .
\label{eq:enhance}
\end{equation}
In Fig.\ 5 this approximate result is presented and
compared with the more exact treatment derived in the previous section,
and with the previous estimate of Lyth \cite{Lyth}.
It is seen that this simple and powerful approximate result is about
$30\%$ low for $\epsilon_{\rm in} \sim 0.15\pi$, and even better at smaller
$\epsilon_{\rm in}$.

\section{Discussion and conclusion}

To describe the whole history of the axion field
and to compute the density fluctuations in the field, one begins with
an assumed value for the mean misalignment angle  $\overline{\theta_{\rm in}}$,
and an assumed distribution about the mean for
$\delta\theta_{\rm in}$, and then
uses the formulation presented here to stretch the fluctuation spectrum
during the time $t\sim\tilde{t}$.
We have obtained a very simple expression for the evolution of axionic
density fluctuations in the entire region
$\overline{\theta_{\rm in}} \in [-\pi,\pi]$,
valid for $\delta\theta/\overline{\theta} \ll 1$ initially.
By using a new model potential,
we have been able to avoid the harmonic and adiabatic approximations,
and still solve the field equations exactly.
Our effective potential has the
twin features of periodicity, and a power--law
temperature dependence of the axion mass (with arbitrary power)
as suggested by the finite temperature instanton approximation \cite{GPY}.

With our model we have extended the heuristic
treatment of ref.\ \cite{Lyth} away from $\theta\sim \pm\pi$.
In ref.\ \cite{Lyth} an estimate of the oscillation period
was used to estimate density fluctuations. Oscillation period,
however, is an ill--defined concept around
the time when $m_{\rm a}\sim 3H$ and damping remains significant; and
is ill--defined
at any time for large initial angles where anharmonic effects are considerable.
Nevertheless, good agreement between our result and
the heuristic result is apparent
in Fig.\ 5 (for the $n=4$ power--law, which is an extrapolation of perturbative
high-temperature instanton effects down to $\tilde{T}$).
The anharmonicity amplifies density \mbox{f}luctuations,
but only significantly so for relatively large initial misalignment angles.
The enhancement factor we find is
$\sim$ (2,3,4,13) for $\theta_{\rm in}\sim (0.85,0.90,0.95,0.99)\times\pi$.

There are possible complications to the axion
evolutionary picture presented here.
These include
\\(i) possible entropy production \cite{entropy}
	and/or nonadiabatic axion mass evolution \cite{ring}
	during the QCD phase transition at $T \sim \Lambda_{\mbox{\scriptsize QCD}}$;
\\(ii) possible damping of the axion oscillations due to interactions with
other
	fields;
\\(iii) possible axion ``trapping" within quark or axion nuggets
	\cite{Mark}, in which the high-temperature phase survives.
\\Let us briefly comment on each of these possibilities.

The density
fluctuations calculated in this paper
are unaffected by entropy production or non--adiabatic mass change during
the QCD phase transition,
provided the axion field is already in the harmonic regime by that time.
This is because Eq.\ (\ref{eq:eom}) stays linear for any temperature
dependence of the axion mass, and the density spectrum is not distorted by
evolution via a linear field equation, as discussed earlier.
{}From the perturbative dilute instanton approximation, Turner has extracted an
estimate \cite{Turner} for the temperature at mass turn--on of
\begin{equation}
\tilde{T}\simeq 0.8\left(\frac{f_{\rm a}/N}{10^{12}\mbox{ GeV}}\right)^{-0.175}
\mbox{GeV}.
\label{eq:star}
\end{equation}
This temperature exceeds $\Lambda_{\mbox{\scriptsize QCD}} \sim$ 100--300 MeV
as long as $f_{\rm a}/N \stackrel{<}{\sim} 3\times10^{14}$~GeV,
or $m_{\rm a} \stackrel{>}{\sim} 2\times 10^{-8}$~eV.
The stronger bound $f_{\rm a} < 10^{12}$~GeV, or $m_{\rm a}>10^{-5}$~eV, that
results from
avoiding overclosure of the Universe with axions
then provides a safety margin here,
although models with hybrid inflation \cite{delta}
or with considerable entropy generation after axion
production \cite{parallel} allow for some relaxation of the overclosure bound.

The interaction of the axion field
with the thermal gluon field provides a damping term \cite{McLerran}
in the equation of motion, Eq.\ (\ref{eq:eom}).  However,
this damping is insignificant at mass turn-on and afterwards if
$f_{\rm a}/N\stackrel{>}{\sim}10^{10}$~GeV, or equivalently,
$m_{\rm a}<0.6\times 10^{-3}$~eV.
As mentioned in the introduction, axions more massive than $10^{-3}$~eV
are disfavored by astrophysical constraints.
If all uncertainties in the astrophysical lower bound on $f_{\rm a}$ conspire,
the $f_{\rm a}$--bound is lowered a bit to $f_{\rm a}>2\times 10^9$~GeV
\cite{Raffelt}.
In some models including non-standard astrophysics \cite{Ellis},
the bound relaxes to $f_{\rm a}>2\times 10^8$~GeV.
Thus it is possible that a $10^{-3}$--$10^{-1}$~eV axion might exist,
in which case interaction--damping becomes important.

In the scenario where axions become trapped in nuggets of high
temperature phase, the low momentum axion
modes which concern us here do not exist.
The axion density fluctuation spectrum, basically flat in inflationary
or stochastic models, is
transformed into a $k^{-2}$ spectrum within a limited
$k$-interval determined by the horizon size at the QCD phase transition,
much too small to affect structure formation \cite{Mark}.
Clearly, in this paper we must assume that
axion trapping does not take place to any significant degree.

The occurrence of large density fluctuations resulting from large initial
field misalignments is possibly impacted by density--fluctuation bounds,
e.g.\ COBE measurements.
Evolution of the spectrum with the formalism presented in this paper
excludes regions in the $m_{\rm a}$--$\theta_{\rm in}$ plane.
Since enhancement of density fluctuations is greatest at large $\theta$,
and since our results at large $\theta$
coincide with those of ref.\ \cite{Lyth},
our exclusions are the same as those of Lyth \cite{Lyth}.
Given a model for the initial spectrum
$\theta_{\rm in}$, or equivalently, for the initial field fluctuation
spectrum $\frac{\delta\theta_{\rm in}}{\overline{\theta_{\rm in}}}$,
these results may exclude particular models.

In summary,
by solving the field equation exactly, we
have provided a tool to
evolve the amplitude fluctuations of the axion field.
Our method is valid for any
scalar field zero--mode (i.e. gradients are neglected)
with temperature--dependent mass,
provided background interactions may be ignored.
The calculation was carried out for a radiation dominated RW metric.
However, the generalization to curvature dominated,
matter dominated, or other RW spacetimes is straightforward.
Our evolution algorithm is ready and able to turn an initial fluctuation
spectrum into a final density contrast.
As the presently unknown initial
fluctuation spectrum becomes more developed theoretically,
this algorithm will become more useful.
Furthermore, recent work on the possible existence of
axion miniclusters \cite{x,thesis,KT} may find a complement in this work.

\vspace{3cm}
{\LARGE\bf Figure Captions}\\
\vspace{0.5cm}

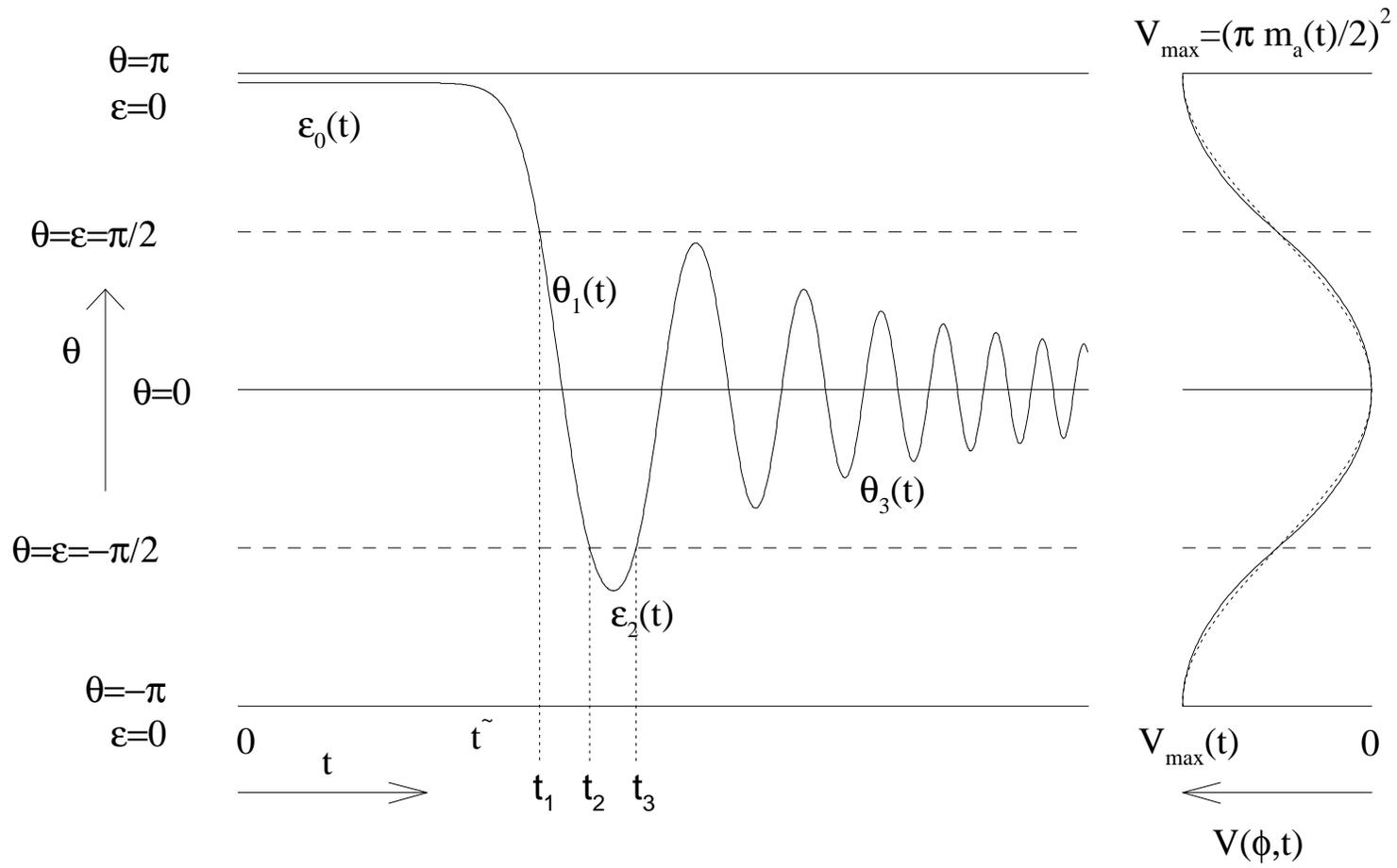
{\bf Figure 1:} A comparison of our model potential (solid line)
with the cosine potential (dashed line); and a
schematic drawing of the field oscillations starting from
a large misalignment angle and evolving through the convex part of the
potential, overshooting it once. The scale on the left illustrates the
conventions we used to relate the functions $\epsilon(z)$ and
$\theta(z)$ to each other. The height of the potential is a function
of time/temperature, but the shape is not.
\vspace{3mm}

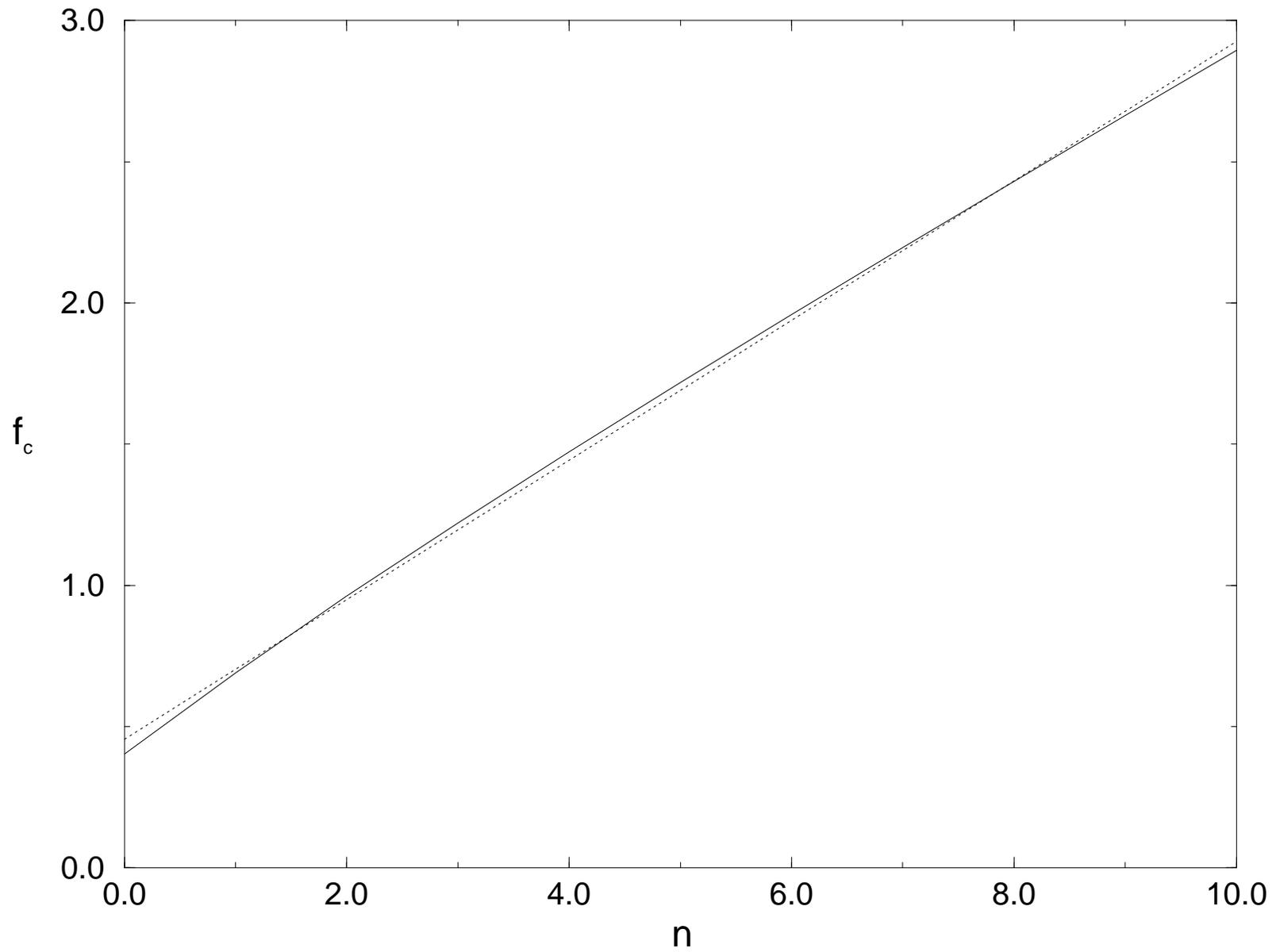
{\bf Figure 2:}
The multiplicative correction factor $f_c(n)$ given in
Eq.\ (\ref{eq:f_c})
for a purely harmonic effective potential with mass turn-on according
to Eq.\ (\ref{eq:m(T)}). The dashed line is the linear fit which has
been extracted numerically by Turner \cite{Turner}.  \vspace{3mm}

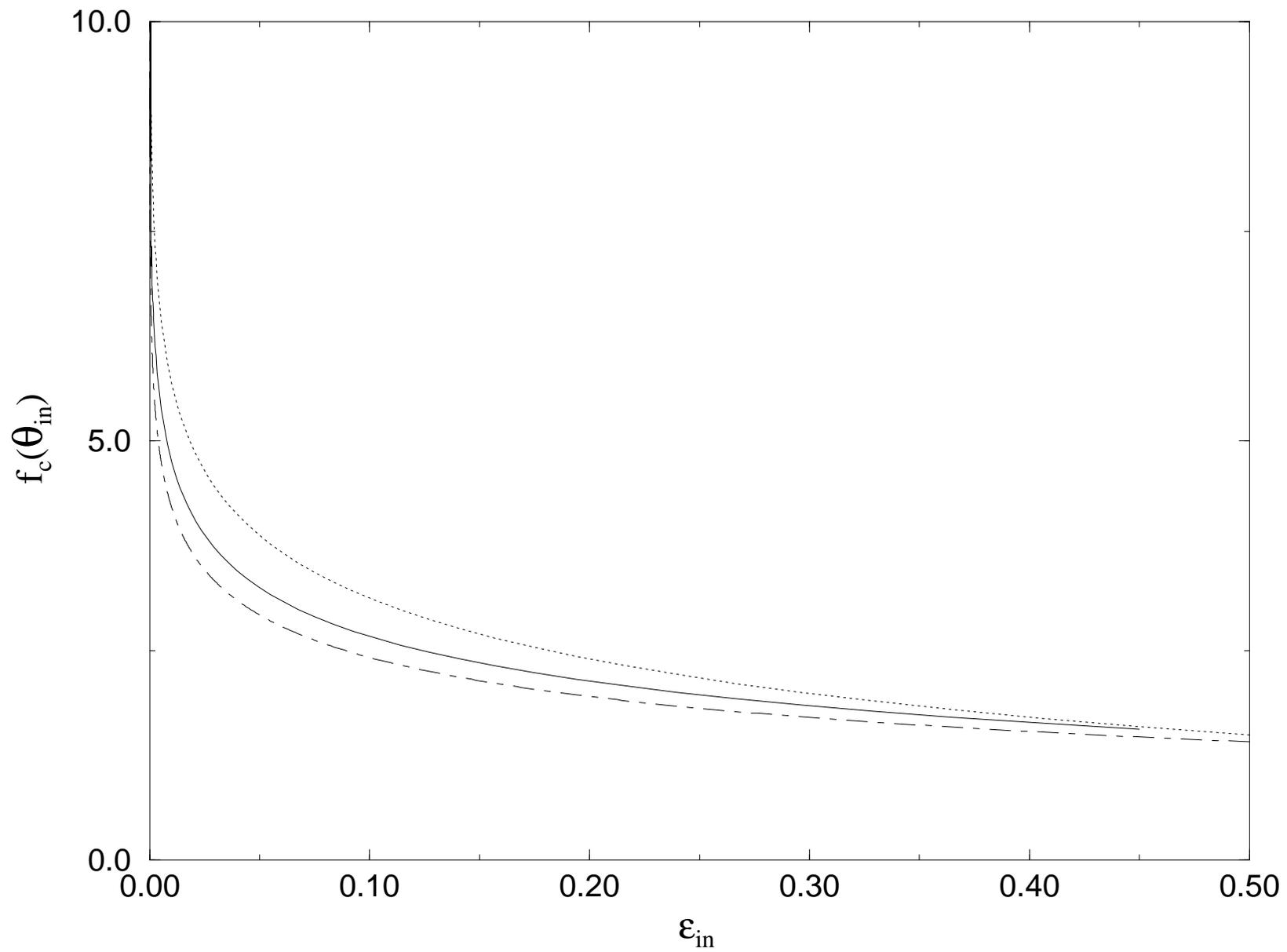
{\bf Figure 3:} The correction factor $f_c(\theta)$,  for $n=0,3.7$ and 8,
to the axionic energy density of Eq.\ (\ref{eq:andens})
predicted by the adiabatic and harmonic approximation.  \vspace{3mm}

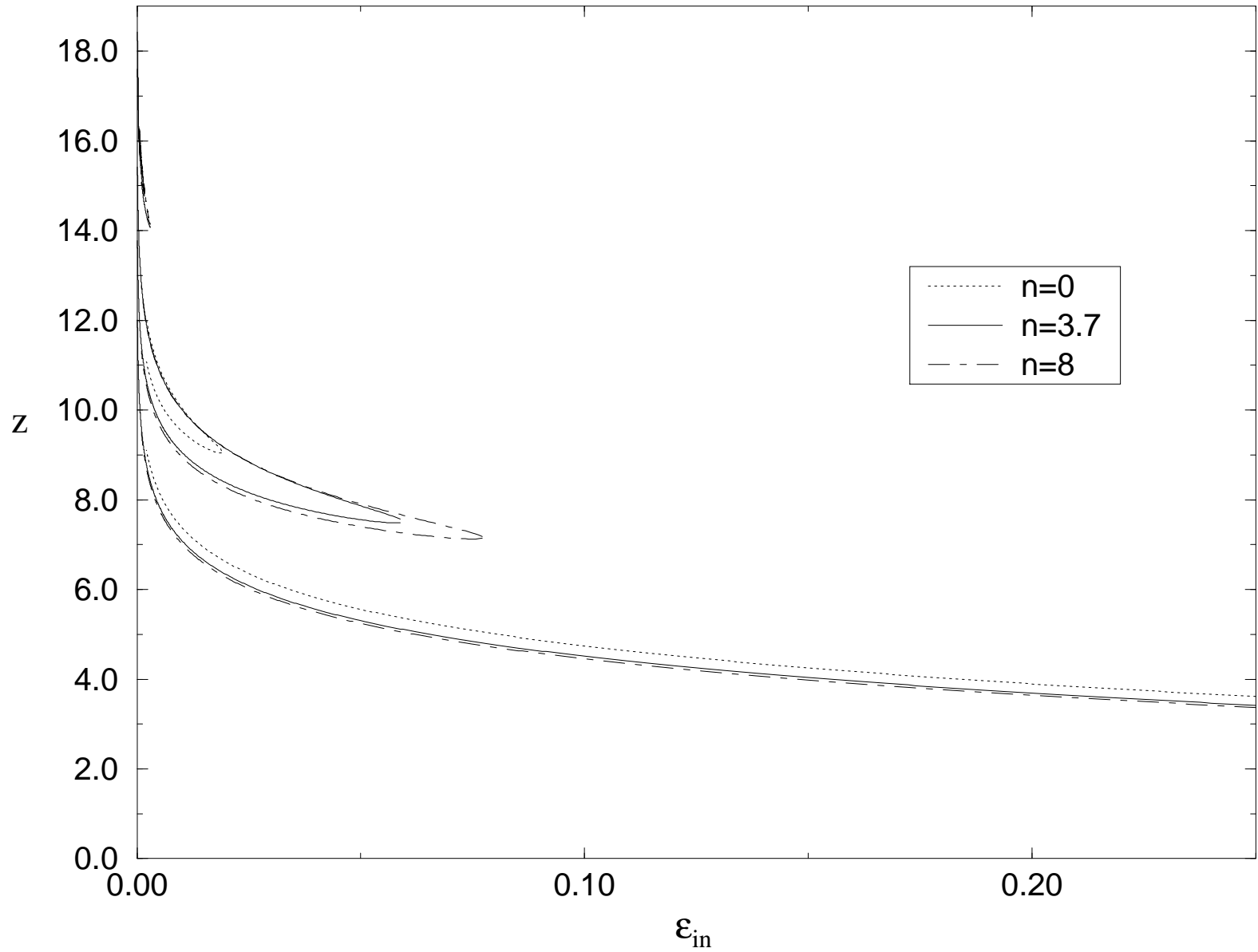
{\bf Figure 4:} The ``exact" $z_1$, $z_2$, $z_3$, $z_4$, and $z_5$
from the transcendental equations with Bessel functions, vs.\
$\overline{\epsilon_{\rm in}}$ for $n=0,3.7$ and 8.
The near equality of $z_2$ and $z_3$ shows that the time spent in the overshoot
solution $\epsilon_2$ is small, especially since for $z\gg1$ the $z$-axis
represents a very stretched $t$-scale, suggesting that truncation of the
solutions with $\theta_1$ is a good approximation.
Note that the values of $z_1$, $z_2$, etc.\,
hardly depend on the choice of $n$, whereas
$t_m=(\frac{n+2}{3} z_m)^{\frac{2}{n+2}}$ values obviously do. The higher is
$n$, the earlier is the onset of oscillations, and the higher is the chance to
overshoot the convex part of the effective axion potential. Using the
asymptotic approximation to calculate $z_1$ gives a graph undistinguishable
from this Figure. \vspace{3mm}

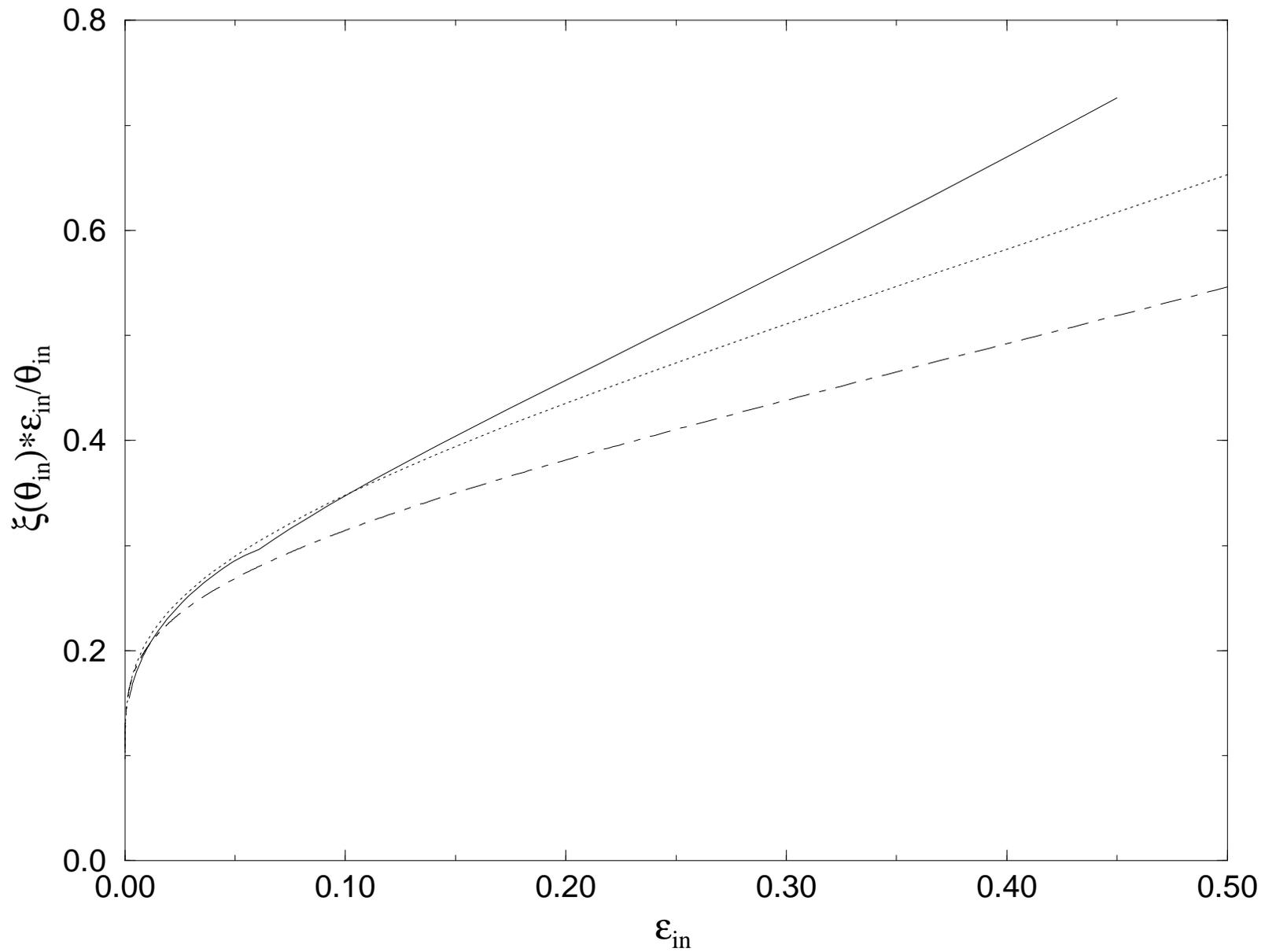
{\bf Figure 5:} $\xi(\overline{\theta_{\rm in}})\frac{\pi-\overline{\theta_{\rm
in}}}
{\overline{\theta_{\rm in}}}$ versus $\overline{\epsilon_{\rm in}}$, with
$n=4$.
The enhancement factor for the density--contrast is $\frac{1}{2}\xi$.
The dot-dashed line is just the result in
asymptotic approximation, given by Eq.\ (\ref{eq:xi2}).
The solid line is the result
using the exact solutions for the coefficients
in Eqs.\ (\ref{eq:00}) and (\ref{eq:ELLIPSE}) and in their successor equations.
Here we considered the
asymptotic solution only up to $\theta_1(z)$, but followed the exact numeric
calculations up to $\theta_3(z)$.  Nevertheless,
the agreement is fantastic, and shows that
repeated reconnecting of the successive solutions is unnecessary.
The discontinuity in the first derivative of the exact
$\xi(\overline{\theta_{\rm in}})$
at $\epsilon_{\rm in}\sim 0.06$ is the result of the field overshooting the
convex region and causing additional enhancement
when $\epsilon_{\rm in}\stackrel{<}{\sim} 0.06$.
The dotted line is a previous estimate made by Lyth \cite{Lyth}. \vspace{3mm}

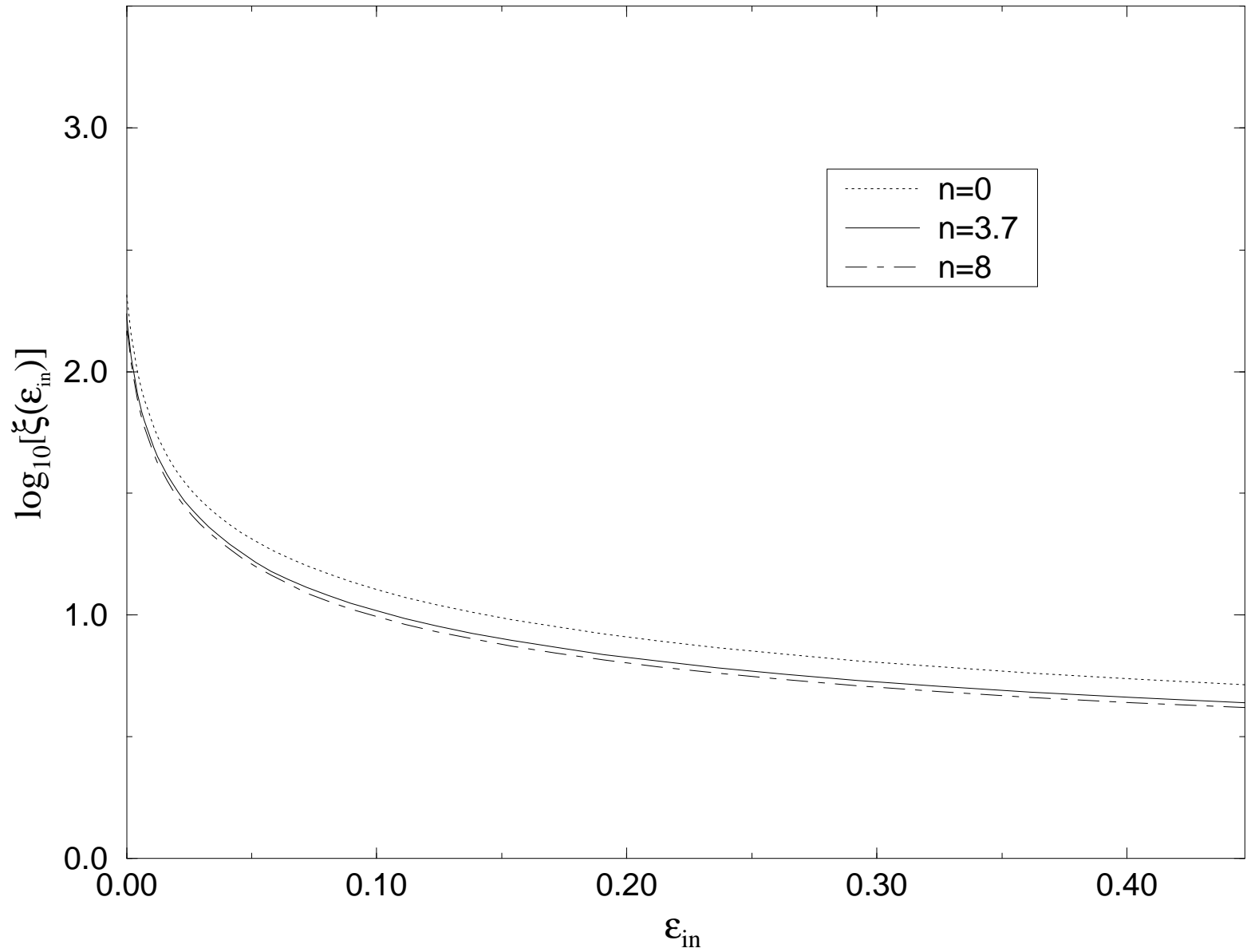
{\bf Figure 6:} $\xi(\overline{\theta_{\rm in}})$ versus
$\overline{\epsilon_{\rm in}}$ for various values of $n$.
We see that the enhancement depends much stronger on the initial
misalignment angle than on the value of $n$. The values here need to be
compared with $\xi(\theta)\equiv 2$ in the harmonic approximation/regime.


\newpage

\begin{thebibliography}{99}
\bibitem{Turner} M.\ S.\ Turner, Phys.\ Rev.\ D {\bf 33},\ 889 (1986)
\bibitem{Lyth} D.\ H.\ Lyth, Phys.\ Rev.\ D {\bf 45},\ 3394 (1992)
\bibitem{PQ} R.\ Peccei and H.\ Quinn, Phys.\ Rev.\ Lett.\ {\bf 38},\ 1440
(1977);\\
             S.\ Weinberg, {\em ibid.}\ {\bf 40},\ 223 (1978);\\
             F.\ Wilczek, {\em ibid.}\ {\bf 53},\ 43 (1981)
\bibitem{delta} A.\ Linde, Phys.\ Lett.\ B {\bf 259},\ 38 (1991)
\bibitem{GPY} D.\ Gross, R.\ Pisarski, and L.\ Yaffe, Rev.\ Mod.\ Phys.\
             {\bf 53},\ 43 (1981)
\bibitem{tilde} In an inflationary scenario,
if the Hubble parameter satisfies $H \ll f_{\rm a}\,$ at the
time when cosmologically relevant scales leave the
horizon, then $\delta\theta/\overline{\theta}\ll 1$ follows, as does a
reheat temperature below the PQ symmetry, which thereby avoids
axionic string production.  These results are presented in
A.\ Vilenkin and L.\ H.\ Ford, Phys.\ Rev.\ D {\bf 25},\ 1231 (1982);\\
     A.\ D.\ Linde, Phys.\ Lett.\ {\bf 116B},\ 335 (1982);\\
     A.\ A.\ Starobinsky, {\em ibid.}\ {\bf 117B},\ 175 (1982)
\bibitem{KT} E. Kolb and I. Tkachev,
	Phys.\ Rev.\ Lett.\ {\bf 71}, 3051 (1993), and
	preprints FERMILAB--PUB--93/335--A, FERMILAB-PUB-94-055-A.
\bibitem{Coleman} C.\ Callan, R.\ F.\ Dashen, and D.\ J.\ Gross, Phys.\ Rev.\ D
             {\bf 17},\ 2717 (1978)
\bibitem{BarTye} W. Bardeen and H. Tye, Phys. Lett. {\bf 74B}, 229 (1978);
	H.--Y. Cheng, Phys. Rept. {\bf158}, 1 (1988).
\bibitem{axgen}
	M. S. Turner, Phys. Rept. {\bf 197}, 67 (1990);
	G.\ Raffelt, Phys.\ Rep.\ {\bf 198},\ 1 (1990).
\bibitem{Oxford} N.\ J.\ Dowrick and N.\ A.\ McDougall, Phys.\ Rev.\ D
              {\bf 40},\ 3486 (1989)
\bibitem{classical axion} J.\ Preskill, M.\ Wise, and F.\ Wilczek, Phys.\
Lett.\
                                     {\bf 120B},\ 127 (1983);\\
        L.\ Abbott and P.\ Sikivie, {\em ibid.}\ {\bf 120B},\ 133 (1983)
\bibitem{Dine} M.\ Dine and W.\ Fischler, Phys.\ Lett.\ {\bf 120B},\ 137 (1983)
\bibitem{Pi} A.\ H.\ Guth and So-Young Pi, Phys.\ Rev.\ D {\bf 32},\ 1899
(1985)
\bibitem{box} Y.\ Nambu and M.\ Sasaki, Phys.\ Rev.\ D {\bf 42},\ 3918 (1990)
present a quantum mechanical description of the growth of density
fluctuations, valid in the limit
$m_{\rm a}^2 \gg \frac{k^2}{R^2} \gg H^2$.
These limits are appropriate for our scenario, and for
structure formation.  However, their formalism calculates
density fluctuations that grow due to
particle creation in the expanding RW background, relevant
only at times long after $\tilde{t}$.
The theory is linearized with gradient terms included,
and so is solved with Fourier analysis.
Thus, an estimate of the spatial scales
(at late times the field gradients eventually reenter the horizon)
as well as the amplitudes
of the fluctuations is possible, given a model for the initial field
distribution.  In contrast, the analysis which we will develop
does not incorporate gradient terms; but it
does follow the zero--mode evolution through the critical
period $t\sim\tilde{t}$, for which there
exists neither a quantum desciption nor a valid analytic classical
description up to now.  Recent work by
D.\ Boyanovsky and R.\ Holman
shows that the quantum corrections to the classical
equation of motion which we follow through the
$t\sim\tilde{t}$ epoch are small and negligible
(private communication with R.\ Holman), further validating our
classical approach.
\bibitem{KoTur} E. W. Kolb and M. S. Turner, \underline{The Early Universe},
	 Addison--Wesley pub., 1990.
\bibitem{thesis} K.\ Strobl, {\em The Probability Spectrum of Axion
     Field Properties in a Non-Inflationary Universe}, Master of Science
	thesis, Vanderbilt University, Nashville, TN, (1992).
\bibitem{entropy} P. Steinhardt and M. S. Turner, Phys. Lett. {\bf 129B}, 51
	(1983).
\bibitem{ring} T.\ DeGrand, T.\ W.\ Kephart, and T.\ J.\ Weiler, Phys.\ Rev.\ D
                                            {\bf 33},\ 910 (1986);
         W.\ Unruh and R.\ Wald, {\em ibid.}\ {\bf 32},\ 831 (1985)
\bibitem{McLerran} L.\ McLerran, E.\ Mottola, and M.\ Shaposhnikov, Phys.\
Rev.\
                                                      D {\bf 43},\ 2027 (1990)
\bibitem{Mark} M.\ Hindmarsh, Phys.\ Rev.\ D {\bf 45},\ 1130 (1992)
\bibitem{parallel} G.\ Lazarides, R.\ K.\ Schaeffer, D.\ Seckel, and Q.\ Shafi,
              Nucl.\ Phys.\ B {\bf 346},\ 193 (1990)
\bibitem{Raffelt} second ref.\ of \cite{axgen}.
\bibitem{Ellis} J.\ Ellis and P.\ Salati, Nucl.\ Phys.\ B {\bf 342},\ 317
	(1990);	T.\ Altherr, Z.\ Phys.\ C {\bf 47},\ 559 (1990).
\bibitem{x} C.\ H.\ Hogan and M.\ J.\ Rees,
	Phys.\ Lett.\ B {\bf 205},\ 228 (1988); E.\ Seidel and W.-M.\ Suen,
        Phys.\ Rev.\ Lett.\ {\bf 72},\ 2516 (1994).
\end{thebibliography}
\end{document}